\documentclass[twocolumn,aps,pra,epsfigure,superscriptaddress,showpacs,eqsecnum]{revtex4}
\usepackage{epsfig}
\usepackage{graphics,color}
\usepackage{amssymb}
\usepackage{amsmath}
\usepackage{amsthm}

\newcommand{\nn}{\nonumber \\}
\newcommand{\bra}[1]{\langle{#1}|}
\newcommand{\ket}[1]{|{#1}\rangle}
\newcommand{\braket}[2]{\langle{#1}|{#2}\rangle}
\newcommand{\lr}[1]{\langle{#1}\rangle}
\newcommand{\re}{{\rm Re}}
\newcommand{\sysphase}{\phi}
\newcommand{\dphi}{\aleph}
\newcommand{\feedphase}{\Phi}
\newcommand{\bestest}{\hat\phi}
\newcommand{\Or}{\Theta}
\newcommand{\nm}{\nu}
\newcommand{\action}{s}

\newcommand{\beq}{\begin{equation}}
\newcommand{\eeq}{\end{equation}}
\newcommand{\sq}[1]{\left[ {#1} \right]}
\newcommand{\cu}[1]{\left\{ {#1} \right\}}
\newcommand{\ro}[1]{\left( {#1} \right)}

\newcommand{\erf}[1]{Eq.~(\ref{#1})}

\newcommand{\tail}{tails}

\newtheorem{thm}{Theorem}

\begin{document}
\title{How to perform the most accurate possible phase measurements}

\author{D.~W.~Berry}
\affiliation{Institute for Quantum Computing, University of Waterloo, Waterloo, ON N2L 3G1, Canada}
\affiliation{Centre for Quantum Computer Technology, Department of Physics, Macquarie University, Sydney, 2109, Australia}
\author{B.~L.~Higgins}
\affiliation{Centre for Quantum Dynamics, Griffith University, Brisbane, 4111, Australia}
\author{S.~D.~Bartlett}
\affiliation{School of Physics, University of Sydney, Sydney, 2006, Australia}
\author{M.~W.~Mitchell}
\affiliation{ICFO---Institut de Ciencies Fotoniques, Mediterranean
Technology Park, 08860 Castelldefels (Barcelona), Spain}
\author{G.~J.~Pryde}
\email{G.Pryde@griffith.edu.au}
\affiliation{Centre for Quantum Dynamics, Griffith University, Brisbane, 4111, Australia}
\author{H.~M.~Wiseman}
\email{H.Wiseman@griffith.edu.au}
\affiliation{Centre for Quantum Dynamics, Griffith University, Brisbane, 4111, Australia}

\begin{abstract}
We present the theory of how to achieve phase measurements with the minimum possible variance in ways that are readily implementable with current experimental techniques. Measurements whose statistics have high-frequency fringes, such as those obtained from NOON states, have commensurately high information yield (as quantified by the Fisher information). However this information is also highly ambiguous because it does not distinguish between phases at the same point on different fringes. We provide schemes to eliminate this phase ambiguity in a highly efficient way, providing phase estimates with uncertainty that is within a small constant factor of the Heisenberg limit, the minimum allowed by the laws of quantum mechanics. These techniques apply to NOON state  and multi-pass interferometry, as well as phase measurements in quantum computing. We have reported the experimental implementation of some of these schemes with multi-pass interferometry elsewhere. Here we present the theoretical foundation, and also present some new experimental results. There are three key innovations to the theory in this paper. First, we examine the intrinsic phase properties of the sequence of states (in multiple time modes) via the equivalent two-mode state. Second, we identify the key feature of the equivalent state that enables the optimal scaling of the intrinsic phase uncertainty to be obtained. This enables us to identify appropriate combinations of states to use. The remaining difficulty is that the ideal phase measurements to achieve this intrinic phase uncertainty are often not physically realizable. The third innovation is to solve this problem by using realizable measurements that closely approximate the optimal measurements, enabling the optimal scaling to be preserved. We consider both adaptive and nonadaptive measurement schemes.
\end{abstract}

\pacs{03.65.Ta, 42.50.St, 03.67.-a}

\maketitle

\section{Introduction}
The measurement of phase is an important task in both metrology and quantum computing. The measurement of optical phase is the basis of much precision measurement, whereas the measurement of the phase encoded in a register of qubits is vital to a broad range of quantum algorithms \cite{Kitaev1996,Cleve1998,Nielsen2000}. In optical phase measurement the precision is usually bound by the standard quantum limit (SQL), where the phase uncertainty is $\Or(N^{-1/2})$ in the number of resources $N$ \footnote{The notation $\Or$ means the asymptotic scaling, as opposed to the notation $O$ which means an upper bound on the asymptotic scaling.}. On the other hand, the fundamental limit imposed by quantum mechanics is $\Or(N^{-1})$ \cite{Caves,Yurke}, often called the Heisenberg limit. There have been many proposals to approach this limit. Ref.\ \cite{Caves} proposed using squeezed states in one port of an interferometer, as well as homodyne measurements, to beat the SQL. Measurements of this type have been experimentally demonstrated \cite{Xiao, Grangier,Goda}. Another type of nonclassical state that has been proposed \cite{Sanders1989,Bollinger1996} and experimentally demonstrated \cite{Lee2002, Rarity1990, Fonseca1999, Edamatsu2002, Walther2004, Mitchell2004, Eisenberg2005, Leibfried2005, Sun2006, Nagata2007, Resch2007} is the NOON state. These provide the maximum phase resolution for a given photon number, although they have the problem that they do not directly provide a unique estimate of the phase.

In quantum computing, the phase, corresponding to the eigenvalue of an operator, can be estimated using Kitaev's algorithm \cite{Kitaev1996}, or the quantum phase estimation algorithm (QPEA) \cite{Cleve1998,Nielsen2000}. The QPEA is based upon applying the inverse quantum Fourier transform (QFT) \cite{Coppersmith,Shor1994}. The inverse QFT, followed by a computational basis measurement, can be applied using just local measurements and control, without requiring entangling gates \cite{Griffiths96}. That simplification allows the phase measurement to be achieved optically, using only linear optics, photodetectors, and electronic feedback onto phase modulators \cite{Higgins07}. The optical implementation could use a succession of NOON states, or a succession of multiple passes of single photons (as was used in Ref.\ \cite{Higgins07}). Using NOON states, or multiple passes, results in high phase sensitivity, but an ambiguous phase estimate. The role of the QPEA is to resolve this ambiguity.

The minimum uncertainty for measurements of a phase shift is $\Or(N^{-1})$ in terms of the total number of applications of that phase shift, $N$, regardless of whether those applications are applied in series or in parallel \cite{vanDam2007}. In quantum optical interferometry, $N$ is the maximum total number of passes of photons through the phase shift. In this formalism, a single pass of an $N$-photon NOON state and $N$ passes of a single photon are regarded as the same number of resources. This is convenient because it enables physical systems that give mathematically identical results to be treated within a unified mathematical formalism. We emphasize that in practice, although these resources are mathematically identical, they are not physically identical, and will be useful in different situations. In particular, the resources for NOON states are used in parallel, which means that they are used within a short space of time. This is needed for measurements where there is a stringent time limit to the measurement, for example due to fluctuation of the phase to be measured, or decoherence of the physical system. In contrast, $N$ passes of a single photon are using resources in series; i.e.\ the applications of the phase shift are sequential. This is useful for measurements of a fixed phase, where there is not an intrisic time limit, but it is required to measure the phase with minimum energy passing through the sample.

In this paper we examine the general problem of how to obtain the most accurate possible phase estimates, by efficiently eliminating ambiguities. This theory applies to general phase measurements in optics (with NOON states or multiple passes) and quantum computation, though for clarity we will primarily present the discussion in terms of NOON states. The problem with simply using the QPEA to eliminate phase ambiguities is that it produces a probability distribution with large \tail, which means that the standard deviation is $\Or(N^{-1/2})$, well short of the Heisenberg limit of $\Or(N^{-1})$. A method of overcoming this problem was presented in Ref.\ \cite{Higgins07}, which used an adaptive scheme to achieve $\Or(N^{-1})$ scaling.
This work was further expanded in Ref.\ \cite{Higgins08}, which proved analytically that scaling at the Heisenberg limit can be achieved without needing adaptive measurements. References \cite{Higgins07,Higgins08} demonstrated these schemes experimentally, using multiple passes of single photons. An alternative scheme based on adapting the size of the NOON states was proposed in Ref.\ \cite{Mitchell2005}. A method of eliminating phase ambiguities in the context of quantum metrology was provided in Ref.\ \cite{Boixo08}.

Here we present the theoretical foundations for Refs.\ \cite{Higgins07,Higgins08}, with further analytical and numerical results, and some new experimental data. First, Sec.\ \ref{sec:limits} explains the limit to the accuracy of phase measurement in more detail. Next, in Sec.\ \ref{sec:theory} the theoretical background to adaptive interferometric measurements and the relation to the QPEA is presented. The equivalent two-mode states when using repeated measurements are presented in Sec.\ \ref{sec:equiv}, and it is shown that (with one exception) they have canonical phase variance scaling as the Heisenberg limit. The adaptive measurements are presented in Sec.\ \ref{sec:adapt}, where the analytical results and numerical results for large $N$ are given. In Sec.\ \ref{sec:reps} it is shown that increasing the number of repetitions leads to a phase uncertainty that still scales as the SQL, rather than the Heisenberg limit. In Sec.\ \ref{sec:simp} the theory for some simplifications to the adaptive scheme is presented. These simplifications include a hybrid scheme as well as a nonadaptive scheme. The approach of adapting the size of the NOON state \cite{Mitchell2005} is given in Sec.\ \ref{sec:size}. Finally, we present the conclusions and a table summarizing the results in Sec.\ \ref{sec:conc}.

\section{Limits to phase measurement}
\label{sec:limits}
The limit to the accuracy of phase measurements can be derived in a simple way from the uncertainty principle for phase \cite{Heitler}
\begin{equation}
\label{eq:unc}
\Delta\phi \Delta n \ge 1/2,
\end{equation}
where the uncertainties are quantified by the square root of the variance. For a single-mode optical field, $n$ is the photon number, and the phase shift is given by the unitary $\exp(i\hat n\phi)$, with $\hat n$ the number operator. More generally, one can consider a phase shift with $\hat n$ being any operator with nonnegative integer eigenvalues. The same uncertainty relation will hold, regardless of the particular physical realization.

The uncertainty principle \eqref{eq:unc} is exact if the variance that is used for the phase is $V_{\rm H} \equiv \mu^{-2}-1$, where $\mu\equiv |\langle e^{i\bestest}\rangle|$, introduced by Holevo~\cite{holevo84}.  The Holevo variance coincides with the usual variance for a narrow distribution peaked well away from the phase cut. Here $\bestest$ is an unbiased estimator of the phase, in the sense that $e^{i\sysphase}=\langle e^{i\bestest}\rangle$. Note that the hat notation is used to indicate a phase estimator, rather than a phase operator. If one has a biased phase estimator, one must  use instead $\mu = \langle\cos(\bestest-\sysphase)\rangle$.

If $n$ is upper bounded by $N$, then the uncertainty in $n$ can never exceed $N/2$. This implies that the phase uncertainty is lower bounded as
\begin{equation}
\Delta\phi \ge 1/N.
\end{equation}
Because this lower bound to the phase uncertainty may be derived from the uncertainty principle for phase, it is usually called the Heisenberg limit. This derivation was presented in terms of the standard deviation for the phase and the mean photon number in Refs.\ \cite{Lane93,Ou96}. As explained in \cite{Lane93,Ou96}, that argument is not rigorous because the uncertainty relation \eqref{eq:unc} is not exact for the usual standard deviation, and the uncertainty in the photon number is not upper bounded by the mean photon number. The above derivation makes the Heisenberg limit rigorous by using the square root of the Holevo variance to obtain an exact uncertainty principle, and using an upper limit on $n$.

The lower bound of $1/N$ on the phase uncertainty is not tight. The exact achievable lower bound, with optimal measurements, is
\begin{equation}
\label{eq:minv}
\Delta\phi_{\rm HL} = \sqrt{V_{\rm H}} = \tan\left(\frac\pi{N+2}\right) \sim \frac{\pi}{N}.
\end{equation}
The asymptotic result was found in the single-mode case in Ref.\ \cite{SumPeg90}, and the exact result was found in Ref.\ \cite{Wiseman97}. This bound in the case of two-mode interferometry, where $N$ is the total number of photons, was found in Refs.\ \cite{Luis96,twomodeprl}. For two-mode interferometry, $n$ is the number of photons passing through the phase shift. An alternative scenario for phase measurement was considered in Ref.\ \cite{vanDam2007}. There it was shown that the same limit holds in a general situation involving $N$ applications of a phase shift to qubits, interspersed with unitaries.

A unified way of defining $N$ that is independent of the physical implementation is 
as follows. The complete measurement scheme, including any feedback, may be represented by preparation of a pure quantum state that depends on a parameter $\phi$, 
followed by measurement. Denote the family of such 
states, parametrized by $\phi$, as $\ket{\psi(\phi)}$. The Fourier transform of this family of states may be taken independently of the basis, and can be denoted $\ket{\tilde\psi(\action)}$. 
Then $N$ can be defined as the minimum size of the interval that supports $\ket{\tilde\psi(\action)}$. See Appendix \ref{sec:meas} for detailed explanations.

Because of the way the Heisenberg limit is derived, there are a number of conditions on its validity:
\begin{enumerate}
\item The error is quantified by the square root of the variance of the difference between the phase estimate and the system phase.
\item $N$ is the total number of applications of the phase shift (the total number of photon passes in quantum optics).
\item There is no {\em a priori} information about the system phase; the prior probability distribution is flat on the interval $[0,2\pi)$. 
\item The variance is evaluated by averaging over all possible system phases, using this flat distribution.
\end{enumerate}
Essentially these conditions require the measurement to be a self-contained measurement of a completely unknown phase. As these are the conditions on the Heisenberg limit, methods to achieve the Heisenberg limit, or scaling as the Heisenberg limit, should satisfy these conditions. For the measurement schemes that we describe here we are careful to ensure that all of these conditions are satisfied. It is possible to derive a similar limit on the measurement accuracy in the phase sensing regime, where it is required to measure small phase shifts \cite{Lee2002}. That limit is also called the Heisenberg limit, though that definition of the Heisenberg limit differs from the definition used here.

The condition that the error is quantified by the square root of the variance is the most stringent condition on the error, because the standard deviation can be used to place bounds on all other commonly used measures of the uncertainty \cite{twomodepra}. On the other hand, it is possible for other measures of uncertainty to give unrealistically small values that are not meaningful. In particular, the reciprocal-peak likelihood can give an uncertainty scaling as $1/\bar n^2$ in the mean photon number \cite{Shapiro89}. That small uncertainty is not meaningful because it does not translate into a correspondingly small standard deviation \cite{Lane93}.

For most other measures of uncertainty, if they are small then (even if the variance for a single measurement is large) results from separate measurements can be combined to yield an overall estimate with small variance. The problem is that this does not necessarily preserve the scaling; a fact which is often ignored in the analysis. If a single measurement uses $N_1$  applications of the phase shift, and there are $M$ repetitions of the measurement, then the total resources are  $N=MN_1$. The variance can then scale no better than  $1/(N_1\sqrt M)=\sqrt M/N$.  Sometimes scaling as $1/(N_1\sqrt M)$ has been referred to as Heisenberg-limited scaling (for a recent example, see Ref.~\cite{MeiHol08}). However, that is not strictly correct because if the number of repetitions $M$ required increases with $N_1$, then the Heisenberg limit of $\Or(1/N)$ scaling will not be obtained, as pointed out recently in Ref.~\cite{Pezze08}. 

Note also that the first condition requires that the error is measured between the actual system phase and the phase estimates. This prevents biased phase estimates giving unrealistically small values of the error. This condition also means that the uncertainty should not be based entirely on the probability distribution for the system phase based on the measurement results. That is, the uncertainty is the spread obtained in the phase estimates from the measurements, rather than the spread in the Bayesian probability distribution obtained from a single measurement.

The second condition is simply that the resources are quantified in the same way as in the proof of the bound. This means that the Heisenberg limit, as defined here, is not beaten by schemes such as those in Refs.\ \cite{Luis04,Caves08}, where the resources are quantified in terms of the number of interacting systems. In these papers the number of applications of the phase shift is equivalent to $\nu t \|H\|$ using their notation (see Appendix \ref{sec:meas}). The variance obtained in Refs.\ \cite{Luis04,Caves08} is not smaller than $\pi/(\nu t \|H\|)$. Note also that, provided the resources are quantified in this way, it is not possible to obtain better scaling than $1/N$ by combining parallel and serial resources. In particular, for optics, $N$ is the number of photon passes, so a state with $n$ photons (the parallel resource) that passes through a phase shift $p$ times (the serial resource) is using total resources $N=np$. This means that it is not possible to obtain better scaling than $1/N$ by, for example, performing the phase measurements we propose with multiple passes of NOON states.

The third condition means that no preexisting information about the phase can be used in the measurement. The fourth condition makes this rigorous by ruling out the possibility of just considering one system phase where the measurement is particularly accurate. This prevents the measurement from implicitly using information about the system phase. In contrast, work on phase measurement often considers the regime of small phase shifts, where the performance of the measurement is considered only for a small range of phases. This approach may be reasonable for states that are close to classical, but breaks down if all the resources are concentrated into a single nonclassical state in order to approach the Heisenberg limit. For example, if all resources are concentrated into measurement with a single NOON state, then accuracy of $\Or(1/N)$ would be obtained only if the phase shift is smaller than $1/N$; otherwise the phase estimate would be ambiguous. That is, for any fixed phase shift, no matter how small, it would not be possible to maintain Heisenberg-limited scaling for arbitrarily large $N$. The phase measurement schemes that we present here show how to eliminate phase ambiguities to achieve accuracy $\Or(1/N)$ without needing \emph{any} initial knowledge of the phase.

\section{Theoretical background}
\label{sec:theory}
\subsection{Interferometric measurements}
\begin{figure}
\centering
\includegraphics[width=0.45\textwidth]{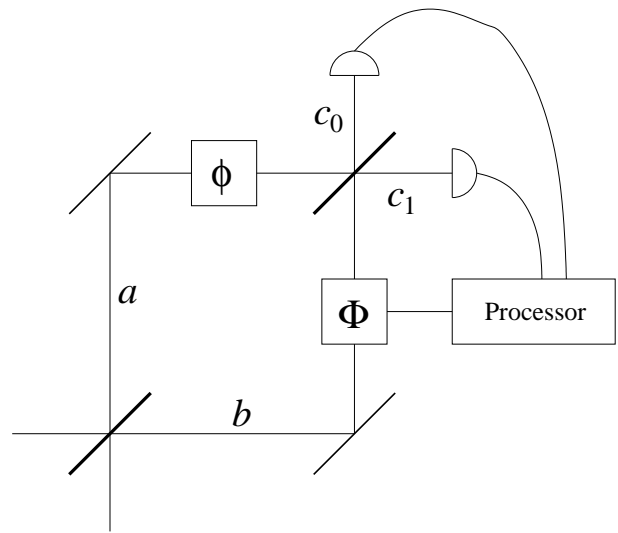}
\caption{The Mach-Zehnder interferometer, with the addition of a controllable phase $\feedphase$ in one arm. The unknown phase to be estimated is $\sysphase$.}
\label{diag0}
\end{figure}

Interferometric measurements are typically considered via the Mach-Zehnder interferometer, as in Fig.\ \ref{diag0}. Two input modes are combined at a beam splitter, after which each of the modes is subjected to a phase shift, and the two modes are recombined at a second beam splitter. The phase shift to be measured, $\sysphase$, is in one arm, and a controllable phase shift, $\feedphase$, may be added in the second arm. The first beam splitter is not necessary for the analysis, and it is more convenient to consider the state in the arms of the interferometer, i.e.\ in modes $a$ and $b$.

The optimal phase measurement, also called the canonical measurement, can be imagined to be performed directly on the modes in the arms, thereby also omitting the second beam-splitter. The phase statistics of such a measurement can be regarded as the intrinsic phase-difference statistics of the state of these two modes.
The joint state of the modes $a$ and $b$ (as shown in Fig.~\ref{diag0}) can be written as a superposition of joint number states. We make the usual assumption that the coefficients in this superposition are positive and real before application of any phase shift.
If the total photon number is fixed at $N$, then the canonical positive operator-valued measure (POVM) is of the form \cite{SandMil95,SandMil97}
\begin{equation}
\label{eq:can2}
F_{\rm can}(\bestest) = \frac 1{2\pi}\ket{\bestest}\bra{\bestest},
\qquad \ket{\bestest} = \sum_{n=0}^{N} e^{in\bestest}\ket{n}\ket{N-n}.
\end{equation}
This POVM gives the probability density for the continuous quantity $\bestest$; that is, the probability of the measurement result being in the interval $A$ is
\begin{equation}
P(\bestest\in A) = \int_{A} {\rm Tr}[F_{\rm can}(\bestest) \rho] d\bestest.
\end{equation} This measurement is applied to modes $a$ and $b$ after application of the phase shift $\sysphase$, but without the phase shift $\feedphase$ (or with $\feedphase$ taken to be zero). The result of the measurement, $\bestest$, is then the estimator for the unknown phase $\sysphase$.

Except in special cases, it is not possible to perform canonical measurements with standard optical equipment (photon counters and linear optical elements such as beam splitters). To approximate canonical measurements, Refs.\ \cite{twomodeprl,twomodepra} use adaptive phase measurements, an idea introduced in Ref.~\cite{Wis95c}. The sequence of detections is used to obtain a Bayesian probability distribution for the phase. The phase $\feedphase$ is then adjusted to minimize the expected variance after detection of the next photon. Note that the time between detections gives no information about the phase. In addition, for fixed total photon number, $N$, the measurement operator for no detection does not alter the state. This means that the time between detections can be ignored in the analysis.

To obtain the probability distribution for the phase, it is convenient to keep a record of the unnormalized system state. The unnormalized system state after $m$ detections, and for system phase $\sysphase$, will be denoted $\ket{\psi(\vec u_{m},\sysphase)}$. Here $\vec u_{m} = (u_1,\ldots, u_m)$ is the vector of $m$ measurement results. The measurement results are taken to be $u_{j} = 0$ or $1$ to indicate that the $j$th photon is detected in mode $c_0$ or mode $c_1$. The annihilation operators at the output of the final beam splitter, $c_0$ and $c_1$, are
\begin{equation}
c_u = [e^{i\sysphase}a + (-1)^u e^{i\feedphase}b]/{\sqrt 2},
\end{equation}
where $a$ and $b$ are the operators for the field at the positions shown in Fig.~\ref{diag0}. After detection, the state is updated as
\begin{equation}
\ket{\psi(\vec u_{m+1},\sysphase)} = \frac{c_{u_{m+1}}}{\sqrt{N-m}}\ket{\psi(\vec u_{m},\sysphase)}.
\end{equation}
The probability for the sequence of measurement results is given by $P(\vec u_{m}|\sysphase)=\braket{\psi(\vec u_{m},\sysphase)}{\psi(\vec u_{m},\sysphase)}$. (The division by $\sqrt{N-m}$ ensures that the probabilities for the different measurement results sum to 1.)

Using Bayes' theorem, the probability for the system phase given the measurement results is
\begin{equation}
P(\sysphase|\vec u_{m})=\frac{P(\sysphase)P(\vec u_{m}|\sysphase)}{P(\vec u_{m})}.
\end{equation}
The initial probability distribution, $P(\sysphase)$, is flat, and the probability $P(\vec u_{m})$ is independent of the phase. Therefore the probability distribution for the system phase is
\begin{equation}
P(\sysphase|\vec u_{m}) \propto P(\vec u_{m}|\sysphase) = \braket{\psi(\vec u_{m},\sysphase)}{\psi(\vec u_{m},\sysphase)}.
\end{equation}
Because the annihilation operators $c_u$ contain the exponential $e^{i\sysphase}$, the unnormalized state and therefore the probability can be expressed in terms of powers of this exponential. This enables the probability distribution to be efficiently represented in terms of the coefficients of the powers of $e^{i\sysphase}$.

The unbiased phase estimate with the smallest variance is $\bestest = \arg \lr {e^{i\phi}}$ \cite{twomodepra}. Here the average is taken over $\phi$, for the Bayesian probability distribution. If the probability distribution $P(\sysphase|\vec u_{m})$ is represented as a Fourier series in $\sysphase$, $\lr {e^{i\phi}}$ is simply the coefficient of the $e^{-i\sysphase}$ term. 
Recall that the Holevo variance in the phase estimates is defined as $V_{\rm H} = \mu^{-2}-1$, where $\mu=|\langle e^{i\hat\phi} \rangle|$.
In order to ensure that the measurement is covariant, we take the initial value of $\Phi$ to be random. Then the value of the sharpness $\mu$ is given by \cite{twomodepra}
\begin{equation}
\label{eq:remu}
\mu = \frac 1{2\pi}\sum_{\vec u_m} \left|\int e^{i\sysphase}P(\vec u_m|\sysphase)d\sysphase \right|.
\end{equation}
See Appendix \ref{sec:det} for a more detailed explanation.

This expression can be rewritten as an average over the sharpnesses of the individual Bayesian probability distributions
\begin{equation}
\label{eq:shav}
\mu = {\rm E}\left[ |\langle e^{i\sysphase}\rangle|\right].
\end{equation}
Here E indicates an expectation value over the measurement results and the initial feedback phase, and the angle brackets indicate an average over $\sysphase$. The way this expression is interpreted is that, for any actual system phase and initial feedback phase, we obtain some measurement results $\vec u_m$ and determine a Bayesian probability distribution for the system phase based on those measurement results.

In Refs.\ \cite{twomodeprl,twomodepra} an algorithm for choosing the feedback phase was introduced, which maximizes the expected sharpness of the Bayesian probability distribution after the next detection. Maximizing the average sharpness, as in Eq.\ \eqref{eq:shav}, minimizes the variance in the phase estimates. Therefore this feedback minimizes the variance in the phase estimates after the next detection. It might be thought that, since the ultimate aim is to minimize the Holevo variance $V_{\rm H}$, the feedback phase should be chosen to minimize the expected Holevo variance of the Bayesian probability distribution after the next detection. However, the variance we want to minimize is that in the phase estimates, not that in the Bayesian probability distributions. If we were minimizing the variance in the Bayesian probability distributions, then we would be minimizing
\beq
 {\rm E} \sq{   |\langle e^{i\sysphase}\rangle|^{-2} -1 }.
\eeq
We actually want to minimize the variance in the phase estimates, which is given by
\beq
V_{\rm H} = \cu{  {\rm E} \sq{ |\langle e^{i\sysphase}\rangle|}}^{-2} -1 .
\eeq
Thus, to minimize $V_{\rm H}$, or indeed any monotonically decreasing function of $\mu$, for the measurement scheme as a whole, one should aim to maximize \erf{eq:remu}, as in Refs.\ \cite{twomodeprl,twomodepra}.

Alternatively, the result that one should aim to maximize \erf{eq:remu} may be shown directly from Eq.\ \eqref{eq:remu}. The feedback phase $\feedphase_{m}$ for detection $m$ is a function of the prior measurement results $u_1,\ldots,u_{m-1}$ and the initial feedback phase $\feedphase_1$. In order to maximize $\mu$ as given by Eq.\ \eqref{eq:remu}, we need to optimize each of these $\feedphase_{m}$ for each measurement record $\vec u_{m-1}$. We can rewrite the sum for $\mu$ in Eq.\ \eqref{eq:remu} as
\begin{equation}
\mu = \frac 1{2\pi}\sum_{\vec u_{m-1}} \sum_{u_m=0}^1\left|\int e^{i\sysphase}P(\vec u_m|\sysphase)d\sysphase \right|.
\end{equation}
Then for each $\vec u_{m-1}$ we have a sum
\begin{equation}
\mu_{\vec u_{m-1}}=\sum_{u_m=0}^1\left|\int e^{i\sysphase}P(\vec u_m|\sysphase)d\sysphase \right|
\end{equation}
that depends only on the feedback phase $\feedphase_m$ for that $\vec u_{m-1}$. Thus the multivariable maximization is reduced to separate maximizations of a single variable for each $\vec u_{m-1}$. As there are only two terms in the sum for $\mu_{\vec u_{m-1}}$, there is an analytical solution for the feedback phase $\feedphase_{m}$ which maximizes this function, as detailed in Ref.~\cite{twomodepra}.

In Ref.\ \cite{twomodepra} it was shown that this approach can, in theory, yield an uncertainty only slightly larger than \eqref{eq:minv} with the optimal input states. The problem is that these states have not yet been produced experimentally. The simplest $N$-photon input state to realize in practice is that with all $N$ photons entering one port of the first beam-splitter in Fig.~\ref{diag0} (with vacuum at the other port). In this case, with the single-pass interferometer design of Fig.~\ref{diag0}, the variance in $\bestest$ can scale no better than $\Or(N^{-1})$ (corresponding to an uncertainty of $\Or(N^{-1/2})$). This limit can not be surpassed regardless of how effective the adaptive measurements are, or even with canonical phase measurements. To beat this scaling it is necessary to use nonclassical multiphoton states, or a variation in the interferometer design that achieves the same end. The nonclassical states with the highest resolution, NOON states \cite{Sanders1989}, give a Holevo variance that is formally infinite, due to the manifold ambiguity in the phase estimate. The same is true if one simply increases the number of passes through the phase shift. Practical solutions to this problem \cite{Higgins07,Higgins08} have come from considering phase measurements in quantum computing, which are discussed in the following subsection.

\subsection{Phase measurements in quantum computing}
In quantum computing, phase estimation is used for finding an eigenvalue of a unitary operator $U$. If an eigenstate, $\ket{u}$, of $U$ is known, the task is to find the corresponding eigenvalue $e^{i\sysphase}$. A control qubit is placed in the state $(\ket 0+\ket 1)/\sqrt 2$, and target qubits are placed in the state $\ket{u}$. A controlled-$U^{2^k}$ operation (i.e., an operation that applies the operator $U^{2^k}$ to a target qubit conditional on the state of the control qubit) then transforms the control qubit into the state $(\ket{0}+e^{i2^k\sysphase}\ket{1})/\sqrt 2$. One approach to finding the phase is the quantum phase estimation algorithm (QPEA) \cite{Cleve1998,Nielsen2000}, in which one performs this procedure on identically prepared qubits for $k=0,\ldots,K$ \cite{Cleve1998}. This yields a state of the form
\begin{equation}
\sum_{y=0}^{2^{K+1}-1}e^{i\sysphase y}\ket y,
\end{equation}
where $\ket y$ is a state on the $K+1$ control qubits. Provided the phase has an exact $K$-bit binary expansion of the form $\sysphase = 2\pi \times 0.a_0\cdots a_K$ (where the $a_j$ are binary digits), performing the inverse QFT yields the state $\ket{a_0\cdots a_K}$, so measurement in the computational basis yields the digits of the phase. If the phase is not of the form $\sysphase = 2\pi \times 0.a_0\cdots a_K$, then accuracy to this number of digits (with success probability $1-\epsilon$) can still be achieved by increasing $K$ to $K'=K+O(\ln(1/\epsilon))$ \cite{Cleve1998}. 

An alternative approach was given by Kitaev \cite{Kitaev1996}. There, rather than using the inverse QFT, separate measurements are used. For each $k$ the number of qubits used is $O(\ln(K/\epsilon))$, which enables $2^k\sysphase$ to be localised in one of 8 subintervals of $[0,2\pi]$ with error probability $\le \epsilon/l$. By combining measurements for different values of $k$, the value of $\phi$ is determined with precision $\pi 2^{-K-2}$ and error probability $\le \epsilon$. In this paper we mostly take inspiration from the QPEA, but in Sec.~\ref{sec:nonadapt} we also use ideas from Kitaev's algorithm. Note that the QPEA was, rather inaccurately, referred to as the Kitaev algorithm in Ref.~\cite{Higgins07}.

Quantum computing algorithms that require phase estimation are for discrete problems, which are typically quantified by the error probability (i.e., the probability that an incorrect answer is given). For such problems, we usually require that the phase error is smaller than a certain amount in order to obtain the correct final answer. If the phase error is larger than this, its magnitude is irrelevant. We contrast this situation to the case in optics,
where phase estimation is typically treated as a continuous problem for which the magnitude of any error is also important. Therefore, the quality of a phase estimation measurement is better quantified by the variance rather than a confidence interval. Also, in quantum computing it is often possible to implement the controlled unitary operation efficiently for arbitrary $k$, so the resources used scale as $K$. In contrast, for optics implementing a multiple of the phase shift requires more resources, so the total resources used scale as $2^K$.

To implement quantum-computing style phase measurements in quantum optics, the multiples of the phase shift can be achieved using either NOON states or multiple passes through a phase shift. The inverse QFT can be implemented by using the simplification of \cite{Griffiths96} which requires only local measurements and control.

Using a single photon as the input to a Mach-Zehnder interferometer, the state in the arms of the interferometer is $(\ket 0+\ket 1)/\sqrt 2$, where $\ket 0$ and $\ket 1$ are the states corresponding to the photon in one arm or the other. Normally the phase shift $\sysphase$ changes the state to $(\ket 0+e^{i\sysphase}\ket 1)/\sqrt 2$. With $\nm$ passes through the phase shift, the state is changed to $(\ket 0+e^{i\nm\sysphase}\ket 1)/\sqrt 2$. The beam splitter then acts as a Hadamard operator, and the photodetectors give a measurement in the computational basis. With a controllable phase of $\feedphase$ in the other arm, the probabilities of the measurement results are
\begin{equation}
P(u|\sysphase) = \frac 12[1+(-1)^u\cos(\nm\sysphase-\feedphase)].
\end{equation}

Alternatively, using a NOON state $(\ket{\nm 0}+\ket{0\nm})/\sqrt 2$ in the arms of the interferometer, the phase shift changes the state to $(e^{i\nm\sysphase}\ket{\nm 0}+\ket{0\nm})/\sqrt 2$. Until detection of all $\nm$ photons, no phase information is obtained, so there is no information on which to base a feedback phase. After detection of all photons, the probability of the sequence of measurement results is
\begin{equation}
P(\vec u_{\nm}|\sysphase) = \frac 12\{1+(-1)^{u_1+\ldots+u_{\nm}}\cos[\nm(\sysphase-\feedphase)]\}.
\end{equation}
This means that the probability distribution for the phase, given the measurement results, only depends on the parity of the measurement results; i.e.\ whether $u_1+\ldots+u_{\nm}$ is odd or even. The phase information obtained is identical to that for a single photon with multiple passes, with the single measurement result $u$ replaced with the parity, and the controllable phase $\feedphase$ replaced with $\nm\feedphase$. From this point on the parity obtained from the NOON measurement will be represented as a single measurement result, $u$.

To apply the phase measurements from quantum computing theory to optics, it is necessary to use multiple time modes. That is, independent single photons are used, or NOON states that are sufficiently spaced apart in time that the photons from the different NOON states can be unambiguously distinguished. The two cases are mathematically equivalent. For brevity, for the remainder of this paper, we present the analysis in terms of NOON states, but the analysis also holds for multiple passes of single photons, as well as phase measurements in quantum computing, because they are mathematically equivalent. The techniques we present can also be applied to Hamiltonian parameter estimation, as discussed in Appendix \ref{sec:meas}, though there is the complication that the range of the parameter is not limited to $[0,2\pi)$. Multiple passes of single photons were used in the experimental demonstration below and in Refs.~\cite{Higgins07,Higgins08}. 

NOON states are often said to provide super-resolution of phase, due to the multiple peaks in the phase distribution obtained in the interval $[0,2\pi]$. If a measurement is performed with just a \emph{single} NOON state, there is a problem in that it is not possible to distinguish which of these peaks corresponds to the phase. This results in the Holevo phase variance being infinite. To obtain a useful phase estimate, the phase information from a NOON state needs to be combined with other phase information.

The QPEA, with the QFT implemented according to the scheme of \cite{Griffiths96}, provides a method to combine the phase information from different NOON states to yield an unambiguous measurement of the phase. Initially the controllable phase $\feedphase$ is set to be zero. The first measurement is taken with a NOON state with $\nm =2^K$. Provided the system phase has an exact $K$-bit binary expression of the form $\sysphase = 2\pi \times 0.a_0\cdots a_K$, this measurement yields the value of $a_K$ as the first measurement result, $u_1$. Given this measurement result, the feedback phase is adjusted by $u_1\pi/2^K$ such that $\sysphase-\feedphase = 2\pi \times 0.a_0\cdots a_{K-1}$. Then the next measurement with $\nm =2^{K-1}$ yields the next digit $a_{K-1}$, and so forth. In this way, all digits $a_k$ are determined with certainty.

This procedure gives a measurement described by a POVM of the form
\begin{equation}
\label{eq:close}
F(\bestest_l) = \ket{\bestest_l}\bra{\bestest_l},
\qquad \ket{\bestest_l} = \frac{1}{\sqrt{N+1}}\sum_{n=0}^{N} e^{in\bestest_l}\ket{n,N-n},
\end{equation}
where $N=2^{K+1}-1$, and $\bestest_l=\pi l/2^K$, for $l=0,\ldots,N$. The state $\ket{n,N-n}$ is labeled by the total number of photons in each arm. That is, $\ket{n,N-n}=\ket n\otimes\ket{N-n}$, where
\begin{equation}
\ket n\equiv\ket{n_0}\otimes\ldots\otimes\ket{n_K},
\end{equation}
and where the $n_k$ are the binary digits of $n$, and similarly for $\ket{N-n}$. 

Since the QPEA gives a phase estimate with $K$ bits of accuracy, at the cost of $2^{K}$ resources, it would seem that it should enable phase variance near the Heisenberg limit. Surprisingly, this is not the case. In fact, it gives a variance above the SQL, as we explain in the next section. 

\section{Equivalent states for multiple time modes}
\label{sec:equiv}

To assist in determining the phase variance yielded by phase measurement algorithms, we introduce the concept of equivalent states. In general, consider two different systems with respective sets of basis states $\{\ket{\chi_n}\}$ and $\{\ket{\xi_n}\}$, such that a phase shift $\sysphase$ adds a factor of $e^{in\sysphase}$ to each basis state. States in the different systems that are identical except for the basis may be regarded as equivalent. That is, $\sum_n \psi_n \ket{\chi_n}$ is equivalent to $\sum_n \psi_n \ket{\xi_n}$. The canonical measurements are identical except for the basis states used, and the distributions obtained for the phase are identical.

In order to analyze a state with multiple time modes, we determine the state with a single time mode that it is equivalent to. We call this the two-mode equivalent state, because it has two spatial modes and just a single time mode. The sequence of $K$ NOON states with the photon number increasing from $1$ to $2^K$ by powers of two is
\begin{equation}
\label{eq:seq}
\ket{\psi_{K}} = \frac 1{2^{(K+1)/2}}\left( \ket{2^K,0}+\ket{0,2^K} \right) \otimes \ldots \otimes \left( \ket{1,0}+\ket{0,1} \right)
\end{equation}
This state is changed under the phase shift to
\begin{align}
\ket{\psi_{K}(\sysphase)} &= \frac 1{2^{(K+1)/2}}\left( e^{i2^K\sysphase}\ket{2^K,0}+\ket{0,2^K} \right) \otimes \nn
& \quad \ldots \otimes \left( e^{i\sysphase}\ket{1,0}+\ket{0,1} \right).
\end{align}
Expanding the product, there is a sum of orthogonal states such as $\ket{2^K,0}\otimes\ket{0,2^{K-1}}\otimes\ldots \otimes\ket{1,0}$. Each state has a coefficient of $e^{in\sysphase}$, where $n$ is the sum of the photon numbers in the first modes from each pair of modes. The state \eqref{eq:seq} is therefore equivalent to, in terms of its phase properties, an equally weighted superposition state with one time mode
\begin{equation}
\label{eq:sup}
\frac 1{2^{(K+1)/2}}\sum_{n=0}^{N} \ket{n,N-n},
\end{equation}
with $N=2^{K+1}-1$.

In terms of these equivalent two-mode states, the POVM for the QPEA can be written as in Eq.\ \eqref{eq:close}, with the states now being the equivalent two-mode states.
This POVM is close to a canonical measurement, except that the possible phase measurement results are restricted to $\bestest_l=\pi l/2^K$. To obtain a canonical measurement, the initial feedback phase $\feedphase_1$ can be taken to be random, rather than zero. For a particular value of this initial feedback phase, the POVM is the same as that in Eq.\ \eqref{eq:close}, except with $\bestest_l=\feedphase_1+\pi l/2^K$. Using random $\feedphase_1$, all $\bestest$ are included in the POVM, so the measurement becomes the canonical phase measurement as in Eq.\ \eqref{eq:can2}.

Using the canonical measurement, the value of $\mu$ for this equal superposition state is $\mu=N/(N+1)= 1-2^{-(K+1)}$. Therefore, the Holevo phase variance is
\begin{equation}
\label{eq:hol}
V_{\rm H} = \frac {1}{(1-2^{-(K+1)})^2} -1 = \frac 2N+\frac 1{N^2}.
\end{equation}
This scales as $1/N$ which is the SQL, rather than the expected Heisenberg limit. To see how this is compatible with the 
precision of the QPEA one must examine the probability distribution for the phase estimate.
Using $\ket{n}$ to denote $\ket{n,N-n}$, this distribution is given by
 \begin{align}
P(\bestest)&=\frac 1{2\pi}\left| \left(\sum_{n=0}^{N} e^{-in\bestest} \bra{n} \right) \left( \frac 1{\sqrt{N+1}}\sum_{n=0}^{N} e^{in\sysphase} \ket{n}\right)\right|^2 \nn &= \frac{\sin^2[(N+1)(\bestest-\sysphase)/2]}{2\pi (N+1)\sin^2[(\bestest-\sysphase)/2]}.
\end{align}
Clearly, this distribution has a peak (around $\sysphase$) with a width $O(N^{-1})$, as expected 
for scaling at the Heisenberg limit. But its variance scales as the SQL because of its high \tail. This is most easily seen using the Collett phase variance $V_{\rm C}\equiv 2(1-\mu)$  \cite{Collett93}. 
The Collett variance coincides with the Holevo variance whenever either is small \cite{twomodepra}.
Using $x=\bestest-\sysphase$ for the phase error, we have
\begin{equation}
V_{\rm C} = \lr{4\sin^2[(\bestest-\sysphase)/2]} = \frac 2{\pi (N+1)}\int_{-\pi}^{\pi} \sin^2(Nx/2) dx.
\end{equation}
That is, the \tail\ are so high that the variance integral is finite only because of the finite range
of the phase. The integral evaluates to $2/(N+1)$, the same SQL scaling as found above for $V_{\rm H}$.  

We have already seen that the QPEA realizes a canonical phase measurement. Therefore if it fails to attain the Heisenberg limit, the fault must lie with the state (\ref{eq:sup}).
One approach is to replace the state \eqref{eq:sup} with a more general state of the form
\begin{equation}
\label{eq:gen}
\sum_{n=0}^{N} \psi_n \ket{n,N-n}.
\end{equation}
Because the measurement is canonical, provided the coefficients $\psi_n$ are optimal the minimum phase uncertainty \eqref{eq:minv} could be obtained.

Unfortunately the minimum phase uncertainty state \cite{SumPeg90,Wiseman97,Luis96,twomodeprl} is not separable between the different time modes, and would therefore be extremely difficult to create. However, it is not necessary for the state to be a minimum phase uncertainty state in order to obtain a phase variance \textit{scaling} as the Heisenberg limit. For a state $\sum_{n=0}^{N}\psi_n\ket{n,N-n}$, the phase variance $V_{\rm C}$ is given by
\begin{equation}\label{var_from_psi}
2(1-|\lr{e^{i\bestest}}|) = \sum_{n=-1}^{N} (\psi_n-\psi_{n+1})^2,
\end{equation}
where $\psi_{-1}$ and $\psi_{N+1}$ are defined to be zero. Provided the state coefficients vary relatively smoothly, the maximum value of $\psi_n$ should be of order $1/\sqrt{N}$, and the successive differences should be of order ${N}^{-3/2}$. This implies that $(\psi_n-\psi_{n+1})^2$ should be of order ${N}^{-3}$, so the variance should scale at the Heisenberg limit of $N^{-2}$. This reasoning shows that the restrictions on the properties of the state are quite weak if one wishes to obtain scaling at the Heisenberg limit.

The equally weighted state does not give phase variance scaling at the Heisenberg limit because there is a large jump, of order $1/\sqrt{N}$, between $\psi_{-1}$ and $\psi_0$ and between $\psi_{N}$ and $\psi_{N+1}$. In order to obtain the required smoothness, one approach is to use multiple copies of the state. The resulting equivalent two-mode state is then the repeated convolution of the original state (in terms of the squared state coefficients). The convolution has the general property of smoothing functions, and therefore can be expected to yield scaling at the Heisenberg limit. In particular we can prove the following theorem.

\begin{thm}
\label{th:mult}
The state
\begin{equation}
\label{eq:mult}
\ket\psi = \frac 1{(N_K+1)^{M/2}}\left( \sum_{n=0}^{N_K}e^{in\phi}\ket{n,N_K-n}\right)^{\otimes M},
\end{equation}
has a canonical phase variance of $\Theta(\ln N_K/N_K^2)$ for $M=2$, and $\Theta(1/N_K^2)$ for $M>2$.
\end{thm}

This theorem means that for fixed $M$ greater than 2, the canonical phase variance scales as $1/N^2$, where $N=N_KM$. The quantity $M$ is the number of copies of the state. The way we achieve a state equivalent to \eqref{eq:mult} is by using $M$ copies of each of the individual NOON states. In applying this theorem, we take $N_K=2^{K+1}-1$, though that is not required for the theorem.
A similar behavior with number of copies was found in Ref.\ \cite{Bagan} for canonical multimode phase estimation (modulo $\pi$) on multiple copies of a squeezed vacuum state, where $M>4$ copies were required to attain the Heisenberg limit. There the optimal number of copies, in terms of minimizing the canonical Holevo variance for a fixed mean photon number, was $8$; in our case it is $3$ (see Fig.\ \ref{fig:Mrun}).

\begin{proof}
Here we give a nonrigorous derivation of the scaling using a continuous approximation. For the rigorous proof of the theorem without using this approximation, see Appendix \ref{app}. The state \eqref{eq:mult} is equivalent to the state
\begin{equation}
\frac 1{(N_K+1)^{M/2}} \sum_{n=0}^{N_K}e^{in\phi}\sqrt{f_M(n)}\ket{n,N_K-n},
\end{equation}
where $f_M(n)$ is the number of combinations of values of $n_1,n_2,\ldots,n_M$ that sum to $n$. This quantity can be regarded as the number of points in a hyperplane perpendicular to a line running between opposite corners of a hypercubic lattice of dimension $M$.

In the continuous approximation, the value of $f_M(n)$ is the area of the cross section of the hypercube, and is equal to $n^{M-1}/(M-1)!$ for $n\le N_K$. The contribution to the variance $V_{\rm C}$ for $n\le N_K$ is then, in the continuous approximation,
\begin{align}
&\frac{1}{(N_K+1)^M}\int_{0}^{N_K} \left(\frac{d\sqrt{f_M(n)}}{dn}\right)^2 dn \nn
&= \frac{1}{(N_K+1)^M} \int_{0}^{N_K} \frac{n^{M-3}}{4[(M-2)!]^2} dn.
\end{align}
There is a clear difference between the cases $M=2$ and $M>2$. For $M>2$, the integral gives an expression proportional to $N_K^{M-2}$, so overall the expression scales as $\Theta(1/N_K^2)$. On the other hand, for $M=2$, the integrand is proportional to $1/n$, rather than of a positive power of $n$ or a constant. The divergence at $n=0$ can be ignored, because the continuous approximation breaks down. However, the integral yields $\ln N_K$ from the upper bound, which means the expression overall scales as $\Theta(\ln N_K/N_K^2)$.

The contribution to $V_{\rm C}$ for $N-N_K\le n\le N$ is the same. For $N_K<n<N-N_K$, we use the fact that $f'_M(n)\le N_K^{M-2}$ and $f_M(n)\ge N_K^{M-1}/(M-1)!$ to find that the integrand is upper bounded by $(M-1)!N_K^{M-3}$. This means that the contribution to $V_{\rm C}$ for $N_k<n<N-N_k$ is $O(1/N_K^2)$ for both $M=2$ and $M>2$. Hence the canonical phase variance, in the continuous approximation, is $\Theta(\ln N_K/N_K^2)$ for $M=2$ and $\Theta(1/N_K^2)$ for $M>2$.
\end{proof}

\section{Adaptive measurements for multiple time modes}
\label{sec:adapt}

\subsection{Deriving the recurrence relation}
The drawback to using these multiple copies of the NOON states is that the canonical measurements can no longer be achieved exactly using adaptive measurements. On the other hand, it is possible to approximate the canonical measurements for larger numbers of repetitions, $M$. The idea is to use a generalization of the adaptive phase measurements for a single time mode. As in that case, at each step the feedback phase $\feedphase$ is chosen to minimize the expected variance after the next detection.

One starts with performing $M$ measurements, one on each of $M$ NOON states, each with photon number $\nm =2^K$ for some integer $K$. Next, measurements are performed on $M$ NOON states with $\nm =2^{K-1}$, and one continues this sequence on NOON states with $\nm =2^k$ for $k=K,K-1,\ldots,1$. This is similar to the optical implementation of the QPEA, except at each stage one uses $M$ NOON states of each size rather than 1. There is a subtlety in using the feedback algorithm of Ref.\ \cite{twomodeprl}, in that when measurements have been performed with NOON states down to size $\nm =2^k$, the phase is only known modulo $2\pi/2^{k}$. To address this, rather than maximizing $|\lr{e^{i\bestest}}|$, the quantity that is maximized is $|\lr{e^{i2^k\bestest}}|$.

Using the same adaptive scheme as before, numerical testing indicates that the phase variances scale as the Heisenberg limit for $M>3$. For small values of $M$ the measurements are a poor approximation of the canonical measurements; for $M=3$ the variance scales as $N^{-3/2}$, and for $M=2$ the variance scales as $N^{-1}$.

The feedback can be achieved in an efficient way when performing measurements on these states. Because the individual NOON states are not entangled with each other, measurements on one do not affect the states of the others, except for the normalization when considering the unnormalized state. Therefore, it is only necessary to keep track of the evolution of the normalization. Recall that the normalization gives the probability distribution for that sequence of detection results, and is proportional to the probability distribution for the phase.

At each stage the probability distribution for the total measurement results is obtained by multiplying the probability for the current measurement result by the preceding probability distribution
\begin{equation}
P(\vec u_m|\sysphase) = P(u_m|\sysphase)P(\vec u_{m-1}|\sysphase).
\end{equation}
The probability distribution for the current measurement result, with a NOON state with $\nm =2^k$, is
\begin{equation}
P(u_m|\sysphase) = \frac 12 \{ 1+(-1)^u \cos[2^k(\sysphase-\feedphase_m)]\}.
\end{equation}
Provided the measurements are made on the largest NOON states first, the overall probability distribution can only contain different powers of $e^{i2^k\phi}$. The probability distribution may therefore be stored using just the coefficients of the different powers. That is, the $p^{(k)}_j$ in the expansion
\begin{equation}
P(\vec u_m|\sysphase) = \sum_j p^{(k)}_j(\vec u_m) e^{ij2^k\sysphase}
\end{equation}
are stored.

The coefficients are updated as
\begin{align}
& p^{(k)}_j(\vec u_m) = \frac 12\left\{ p^{(k)}_j(\vec u_{m-1}) + \frac {(-1)^{u_m}}2 \right. \nn
& \left.\times\left[ 
p^{(k)}_{j-1}(\vec u_{m-1}) e^{-i2^k\feedphase_M}+ p^{(k)}_{j+1}(\vec u_{m-1}) e^{i2^k\feedphase_M} \right]\right\}.
\end{align}
When examining measurements for the next lower value of $\nm$, given by $\nm =2^{k-1}$, the coefficients are expanded out by a factor of two. That is, the probability distribution is written as
\begin{equation}
P(\vec u_m|\sysphase) = \sum_j p^{(k-1)}_j(\vec u_m) e^{ij2^{k-1}\sysphase},
\end{equation}
where $p^{(k-1)}_{2j}(\vec u_m) = p^{(k)}_j(\vec u_m)$ and $p^{(k-1)}_{2j+1}(\vec u_m)=0$.

In addition, it is only necessary to keep track of a limited number of the $p^{(k)}_j(\vec u_m)$. At the end of the measurements, the quantity of importance is $p^{(0)}_1(\vec u_{M(K+1)})$. This can depend only on the $p^{(k)}_j(\vec u_{m})$ for $j$ in the range $-2M,\ldots,2M$, and only these coefficients need be recorded. In fact, only coefficients for $j=0,\ldots,2M$ need be recorded, because those for negative $j$ are the complex conjugate of those for positive $j$.

To illustrate these ideas, consider the first measurement for $\nm=2^k$, where the first few coefficients are
\begin{align}
p_0^{(k)}(\vec u_{m}) &= p_0^{(k+1)}(\vec u_{m}), \nn
p_1^{(k)}(\vec u_{m}) &= 0, \nn
p_2^{(k)}(\vec u_{m}) &= p_1^{(k+1)}(\vec u_{m}), \nn
p_3^{(k)}(\vec u_{m}) &= 0,
\end{align}
where $m=M(K-k)$. After the next detection, the first few coefficients are changed to
\begin{align}
p_0^{(k)}(\vec u_{m+1}) &= p_0^{(k+1)}(\vec u_{m})/2, \nn
p_1^{(k)}(\vec u_{m+1}) &= (-1)^{u_{m+1}}[p_0^{(k+1)}(\vec u_{m})e^{-i2^k\feedphase_{m+1}} \nn & \quad + p_1^{(k+1)}(\vec u_{m})e^{i2^k\feedphase_{m+1}}]/4, \nn
p_2^{(k)}(\vec u_{m+1}) &= p_1^{(k+1)}(\vec u_{m})/2.
\end{align}
The feedback phase is chosen to maximize the average of $|p_1^{(k)}(\vec u_{m+1})|$ over the two values of $u_{m+1}$. For both measurement results,
\begin{equation}
|p_1^{(k)}(\vec u_{m+1})| = \frac 14 |p_0^{(k+1)}(\vec u_{m})e^{-i2^{k+1}\feedphase_{m+1}}+p_1^{(k+1)}(\vec u_{m})|.
\end{equation}
Because $p_0^{(k)}(\vec u_{m})$ is real and positive (the probability distribution is real and positive), the feedback phase that maximizes this average is
\begin{equation}
\label{eq:fed1}
\feedphase_{m+1} = -2^{-(k+1)}\arg [p_1^{(k+1)}(\vec u_{m})].
\end{equation}
We now investigate this solution for various values of $M$.

\subsection{Analytic result for $M=1$}
In the case $M=1$, Eq.\ \eqref{eq:fed1} yields a recurrence relation that can be solved, yielding \footnote{For this solution the $\arg$ function in \eqref{eq:fed1} must be taken to give values in the interval $(-2\pi,0]$, so $\feedphase$ is positive.}
\begin{align}
& p_0^{(k)}(\vec u_{K-k+1}) = 2^{k-K-1}, \nn
& p_1^{(k)}(\vec u_{K-k+1}) = \frac{2^{K-k+1}-1}{2^{2(K-k+1)}} \nn
& \quad\times\exp\left(-i\pi\sum_{l=1}^{K-k+1}u_l/2^{K-k+1-l}\right).
\end{align}
The feedback phase obtained is
\begin{align}
\feedphase_{K-k+1} &= 2\pi\sum_{l=1}^{K-k}u_l/2^{K+1-l} \nn
&= 2\pi\times 0.00\ldots 0u_{K-k}u_{K-k-1}\ldots u_1,
\end{align}
where the expression following the ``$\times$'' on the second line is a binary expansion. For each measurement result, $u_l$, the feedback phase is adjusted by the appropriate fraction of $2\pi$. This feedback is identical to that obtained for the linear optics implementation of the QPEA described in Sec.\ \ref{sec:theory}. In addition, summing over $|p_1^{(k)}(\vec u_{K-k+1})|$ at the end of the measurement gives $\mu=1-2^{-(K+1)}$. This yields exactly the same variance as in Eq.\ \eqref{eq:hol}.

\subsection{Analytic result for $M=2$}
It is also possible to analytically determine the variance for $M=2$. In this case, the values of $p_0$ and $p_1$ after the second measurement for a given $k$ are
\begin{align}
&p_0^{(k)}(\vec u_{m+2}) = p_0^{(k)}(\vec u_{m})/4+\frac{(-1)^{u_{m+1}}}8 \re[p_0^{(k)}(\vec u_{m}) \nn
&\quad\times e^{i2^k(\feedphase_{m+2}-\feedphase_{m+1})}+p_2^{(k)}(\vec u_{m})e^{i2^k(\feedphase_{m+2}+\feedphase_{m+1})}], \nn
&p_1^{(k)}(\vec u_{m+2}) = \frac{(-1)^{u_{m+1}}}8[p_0^{(k)}(\vec u_{m})e^{-i2^k\feedphase_{m+1}}+p_2^{(k)}(\vec u_{m})\nn &\quad\times e^{i2^k\feedphase_{m+1}}]+
\frac{(-1)^{u_{m+2}}}8[p_0^{(k)}(\vec u_{m})e^{-i2^k\feedphase_{m+2}} \nn & \quad+p_2^{(k)}(\vec u_{m})e^{i2^k\feedphase_{m+2}}].
\end{align}
Taking the absolute value of $p_1$ yields
\begin{align}
& p_1^{(k)}(\vec u_{m+2}) = \frac 18 \left| e^{i2^{k-1}(\feedphase_{m+2}-\feedphase_{m+1})} \right. \nn & \left. \quad 
+ (-1)^{u_{m+2}-u_{m+1}}e^{i2^{k-1}(\feedphase_{m+2}-\feedphase_{m+1})}\right| \nn & \quad \times \left| p_0^{(k)}(\vec u_{m}) - p_2^{(k)}(\vec u_{m})e^{i2^k(\feedphase_{m+2}+\feedphase_{m+1})} \right|.
\end{align}
Regardless of whether the feedback phase \eqref{eq:fed1} is assumed for $\feedphase_{m+1}$, or both $\feedphase_{m+1}$ and $\feedphase_{m+2}$ are maximized over, the solution is
\begin{align}
2^k(\feedphase_{m+2}-\feedphase_{m+1}) &= \pi/2, \nn
2^k(\feedphase_{m+2}+\feedphase_{m+1}) &= -\arg[p_2^{(k)}(\vec u_{m})] + \pi/2.
\end{align}
This means that the feedback scheme minimizes the variance two detections in advance, rather than just one. Independent of the detection results, the absolute value obtained is
\begin{equation}
|p_1^{(k)}(\vec u_{m+2})| = \frac {\sqrt 2}8 \sqrt{ [p_0^{(k)}(\vec u_{m})]^2+|p_2^{(k)}(\vec u_{m})|^2}.
\end{equation}
Using this feedback also gives
\begin{equation}
p_0^{(k)}(\vec u_{m+2}) = p_0^{(k)}(\vec u_{m})/4.
\end{equation}

This again yields a recurrence relation that can be solved. The result is
\begin{align}
p_0^{(k)}(\vec u_{m+2}) &=2^{2(k-K-1)}, \nn
|p_1^{(k)}(\vec u_{m+2})| &= 2^{2(k-K-1)} \sqrt{1-2^{k-K-1}}.
\end{align}
At the end of the measurement, summing over the $|p_1^{(0)}(\vec u_{2(K+1)})|$ gives $\mu = \sqrt{1-2^{-(K+1)}}$.
This yields the Holevo variance in this case as
\begin{equation}
V_{\rm H} = \frac 1{2^{K+1}-1} = \frac 2{N}.
\end{equation}
Hence, in this case, the Holevo phase variance is \emph{exactly} $2/N$. This is almost identical to the result for $M=1$, where the variance is approximately $2/N$. This means that, despite the intrinsic state having relatively good phase properties (with variance scaling as $(\ln N)/N^2$), the adaptive measurements are far less accurate than the canonical measurement. Furthermore, because the feedback phases minimize the variance two detections in advance, it seems unlikely that a better feedback scheme would yield significantly more accurate measurements.

\subsection{Numerical results for $M>2$}
For larger values of $M$, more accurate measurements are obtained, but we have not found exact analytical results. It is possible to obtain exact results for smaller values of $K$ simply by systematically calculating the final values of $p_1^{(0)}$ for every possible combination of measurement results. For larger values of $K$ this approach is no longer feasible, but instead the variance can be estimated by generating random system phases, then generating measurement results according to their probability of occurring. There are then two alternative ways of estimating the phase variance. One is to use the mean of the final values of $|p_1^{(0)}|$ as an estimate of $\mu$. The other is to use the variance in the final phase estimates relative to the system phase. Numerically it is found that using the values obtained for $|p_1^{(0)}|$ gives a more accurate estimate of the phase variance.

\begin{figure}
\centering
\includegraphics[width=0.48\textwidth]{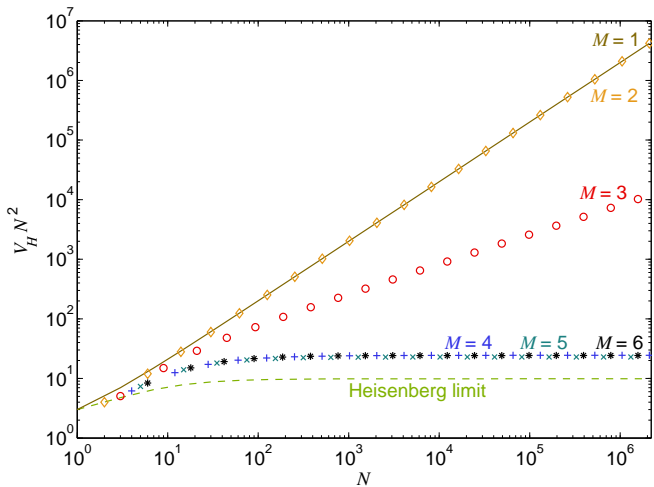}
\caption{The Holevo phase variance multiplied by $N^2$ for $M=1$ to 6. The results for $M=1$ are shown as the solid line, for $M=2$ as the diamonds, for $M=3$ as the circles, for $M=4$ as the pluses, for $M=5$ as the crosses, and for $M=6$ as the asterisks. The Heisenberg limit is also shown as the dashed line for comparison.}
\label{fig:var1}
\end{figure}

The results for $M=1$ to 6 are shown in Fig.\ \ref{fig:var1}. The variances are multiplied by $N^2$, so that the Heisenberg limit appears as a horizontal line. The lines for $M=1$ and 2 are not scaling as the Heisenberg limit, and are indistinguishable on this plot for $N$ above about 10. The variances in these cases scale as $1/N$, and the numerically obtained variances agree with the analytical predictions to within the numerical precision. The lines for $M=4,5$ and 6 are clearly scaling as $1/N^2$. The line for $M=3$ has intermediate scaling, and is consistent with an asymptotic scaling of $1/N^{3/2}$.

\begin{figure}
\centering
\includegraphics[width=0.48\textwidth]{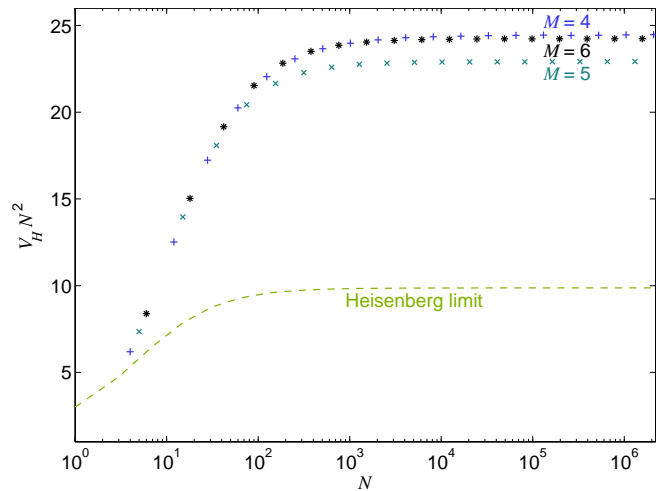}
\caption{The Holevo phase variance multiplied by $N^2$ for $M=4,5$ and 6. The results for $M=4$ are shown as the pluses, for $M=5$ as the crosses, and for $M=6$ as the asterisks. The Heisenberg limit is also shown as the dashed line for comparison.}
\label{fig:var2}
\end{figure}

Although the results for $M=4,5$ and 6 scale as the Heisenberg limit, they have different scaling constants. The variances for these cases, multiplied by $N^2$, are shown on a linear scale in Fig.\ \ref{fig:var2}. The smallest scaling constant of about 23 is obtained for $M=5$.

\section{The effect of increasing the number of repetitions}
\label{sec:reps}
\begin{figure}
\centering
\includegraphics[width=0.48\textwidth]{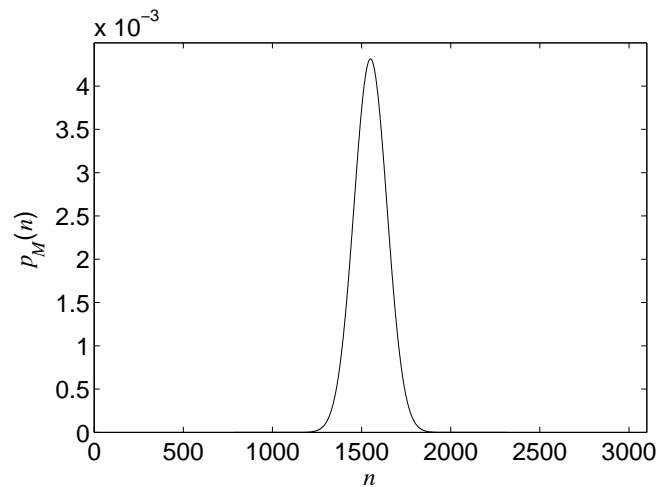}
\caption{The squares of the state coefficients for the equivalent two-mode state for $K=4$ and $M=100$.}
\label{fig:middle}
\end{figure}

In order to obtain higher precision, there are two parameters that can be increased. The value of $K$ can be increased, or the number of repetitions, $M$, can be increased. Scaling at the Heisenberg limit can only be obtained by increasing $K$, not by increasing $M$, because the state coefficients do not gradually increase from zero in the latter case. Instead they are negligible for most values of $n$, and suddenly rise to significant values near the centre (for example, see Fig.\ \ref{fig:middle}).

The result can be shown rigorously using the uncertainty relation (in terms of the variance) $V_{\rm H}(\sysphase)V(n)\ge 1/4$ \cite{holevo84}, where $V(n)$ is the variance for the photon number. To determine the number variance, take $p_M(n)=f_M(n)/(N_K+1)$, and use the recurrence relation (see Appendix \ref{app})
\begin{equation}
\label{eq:magic}
f_{M}(n) = \sum_{k=0}^{N_K} f_{M-1}(n-k).
\end{equation}
This yields the recurrence relation $\lr n_M=\lr n_{M-1}+N_K/2$ (where $\lr n_M=\sum_n n p_M(n)$). This gives $\lr n_M = MN_K/2$. The same approach for $\lr{n^2}_M$ gives the recurrence relation
\begin{equation}
\lr{n^2}_M= \lr{n^2}_{M-1} + N_K(2N_K+1)/6 + N_K\lr n_{M-1}. 
\end{equation}
Solving this gives $\lr{n^2}_M=MN_K(N_K+2)/12+M^2N_K^2/4$, so $V(n) = MN_K(N_K+2)/12$. Using the uncertainty relation, we obtain the lower bound on the phase variance
\begin{equation}
\label{eq:repsbnd}
V_{\rm H}(\sysphase) \ge \frac 3{MN_K(N_K+2)}.
\end{equation}
Thus, with fixed $K$, the variance can only scale as $1/M$, not as $1/M^2$. On the other hand, because the number variance scales as the square of $N_K$, the Heisenberg limit can be achieved when increasing $K$, as we have already seen.

\begin{figure}
\centering
\includegraphics[width=0.48\textwidth]{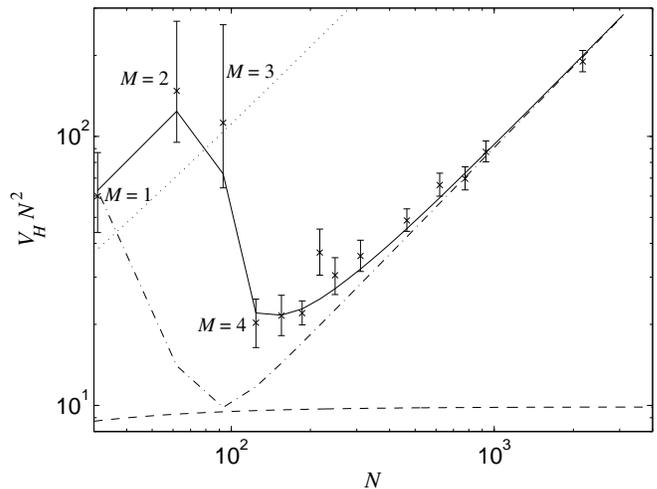}
\caption{The theoretical predictions and experimental results for adaptive measurements with $K=4$ and a range of values of $M$ (given by $M=N/31$). The solid line is the predictions for the adaptive scheme, the dashdotted line is the intrinsic variance, and the crosses and error bars are for the experimental data. The SQL for single passes is shown as the dotted line, and the dashed line is the Heisenberg limit.}
\label{fig:Mrun}
\end{figure}

The predicted phase variance obtained using the adaptive measurement scheme with $K=4$ is plotted as a function of $M$ in Fig.\ \ref{fig:Mrun}. Experimental results using multiple passes are also shown in this figure, and agree with the theoretical predictions. These experimental results were obtained using an apparatus and methodology identical to that described in Ref.\ \cite{Higgins07}. When increasing $M$, the variance at first decreases rapidly, and approaches the Heisenberg limit. Then, as $M$ is increased further, the variance no longer decreases at the same rate as the Heisenberg limit, and instead scales as the SQL for single passes.

\section{Simplifications of the adaptive measurements}
\label{sec:simp}
The adaptive scheme described above is complicated to implement in practice due to the calculation required to determine the feedback phase. This calculation is non-Markovian, in the sense that it depends in a nontrivial way on all the prior measurement results. For this reason it is useful to develop schemes using simplified (Markovian) feedback, or no feedback. Two alternative schemes were proposed and experimentally demonstrated in Ref.\ \cite{Higgins08}. Here we provide further motivation and theoretical analysis of these alternative measurement schemes.

\subsection{Hybrid scheme} \label{sec:Hybrid}
One scheme is based on a simple modification of the QPEA. The QPEA is Markovian, because the \emph{increment} in the feedback phase depends only on the immediately preceding result. As discussed above, the QPEA gives a canonical measurement, but the state is equivalent to an equal superposition, and thus has poor phase properties. It yields a probability distribution for the phase with \tail\ that drop off slowly, and give the dominant contribution to the phase variance. The idea behind hybrid measurements is to use additional ``standard'' measurements (single passes of single photons) to reduce the size of the \tail\ of the distribution. The hybrid measurements are still Markovian, because they do not use any feedback beyond what is used in the QPEA. However, these measurements do not scale at the Heisenberg limit, and instead the variance scales at best as close to $1/N^{3/2}$. In particular, the result is given by the following theorem.

\begin{thm}
\label{th:hyb}
The canonical phase variance of the state
\begin{align}
\ket{\psi_{N_K,M}} &= \frac 1{2^{M/2}\sqrt{N_K+1}}\left(\sum_{n=0}^{N_K} e^{in\phi} \ket{n,N_K-n}\right)\nn
&\quad\otimes\left(\ket{0,1}+e^{i\phi}\ket{1,0}\right)^{\otimes M},
\end{align}
scales as $\Omega(N^{-3/2})$, where $N=N_K+M$.
\end{thm}

Here $\Omega$ is the standard notation for a lower bound on the scaling. In applying this theorem, we take $N_K=2^{K+1}-1$ to be the number of photons used in the QPEA, and $M$ to be the number of photons used in the ``standard'' interferometry.

\begin{proof}
To obtain the equivalent two-mode state for $M$ repetitions of measurements with single photons we apply the recurrence relation \eqref{eq:magic}. Repeatedly applying this recurrence relation gives
\begin{equation}
f_M(n)={M \choose n}.
\end{equation}
This result can also be obtained from the fact that this number is the number of ways of choosing $n$ ones from $M$ bits. Then applying the recurrence relation again with the flat distribution gives
\begin{equation}
f(n)=\sum_{k=0}^{N_K} {M \choose n-k}.
\end{equation}
This gives the probability distribution for the $n$,
\begin{equation}
p(n)=\frac 1{2^M(N_K+1)}\sum_{k=0}^{N_K} {M \choose n-k}.
\end{equation}
Evaluating the moments for this probability distribution gives
\begin{align}
\lr n &= (N_K+M)/2, \nn
\lr{n^2} &= \frac M4 (M+1)+\frac{N_KM}2+\frac {N_K}6(1+2N_K).
\end{align}
This gives the variance in $n$ as $(3M+2N_K+N_K^2)/12$. As above, using the uncertainty relation for the Holevo variance yields
\begin{equation}
V_{\rm H}(\bestest) \ge \frac 3{3M+2N_K+N_K^2}.
\end{equation}
This relation implies that if $N_K \in O(N^{1/2})$, then the SQL is obtained.
To obtain better scaling than the SQL, it is necessary for $N_K$ to increase with $N$ more rapidly than $\sqrt{N}$.

This result on its own does not not show that the intrinsic variance cannot scale better than $N^{-3/2}$, and it is necessary to also examine other features of the equivalent two-mode state. The problem with the equivalent two-mode state is that the state coefficients rapidly increase, then there is a wide flat region (for example, see Fig.\ \ref{fig:shelf}), rather than the gradual increase to a maximum which is required.

\begin{figure}
\centering
\includegraphics[width=0.48\textwidth]{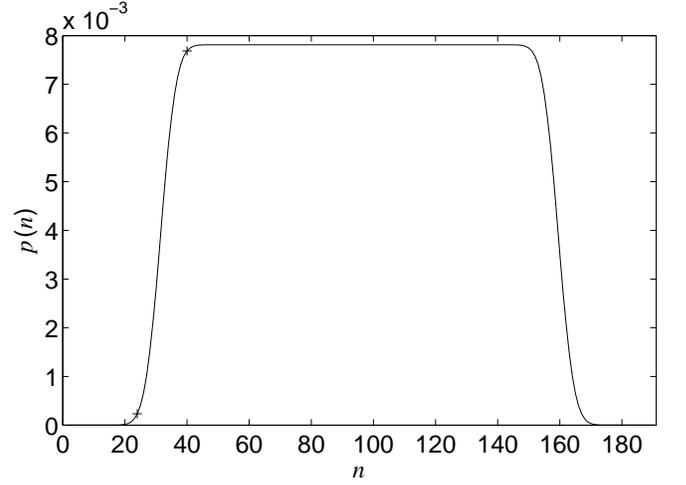}
\caption{The squares of the state coefficients for the equivalent two-mode state for hybrid measurements with $N_K=127$ and $M=64$. The pluses show the positions of $n_-$ and $n_+$.}
\label{fig:shelf}
\end{figure}

To put a bound on the phase variance, consider the value of $p(n)$ for $n=n_-=\lfloor M/2-\sqrt M \rfloor$ and $n=n_+=\lfloor M/2+\sqrt M \rfloor$. These values are chosen as points between which the $p(n)$ vary rapidly (see Fig.\ \ref{fig:shelf}). Because $M<N$, and $N_K$ increases more rapidly than $\sqrt{N}$,
asymptotically we must have $N_K>2\sqrt M$. Using this, and Hoeffding's inequality \cite{Hoeffding}, we obtain 
\begin{align}
p(n_+) &= \frac 1{2^M(N_K+1)}\sum_{k=n_+-N_K}^{n_+} {M \choose k} \nn
&\ge \frac 1{2^M(N_K+1)}\sum_{k=n_-}^{n_+} {M \choose k}, \nn
&> (1-2e^{-2})/(N_K+1)
\end{align}
and
\begin{align}
p(n_-) &= \frac 1{2^M(N_K+1)}\sum_{k=n_--N_K}^{n_-} {M \choose k}. \nn
&\le e^{-2}/(N_K+1)
\end{align}
In addition, it is easily verified that $n_+-n_-\le 4\sqrt{M/3}$, so
\begin{align}
\sum_{n=n_-}^{n_+-1} (\sqrt{p(n+1)}-\sqrt{p(n)})^2 
&\ge \frac{(\sqrt{p(n_+)}-\sqrt{p(n_-)})^2}{n_+-n_-} \nn
&\ge \frac{\left( \sqrt{1-2e^{-2}}-e^{-1}\right)^2}{4 (N_K+1)\sqrt{M/3}} \nn
&= \frac 1{2\kappa(N_K+1)\sqrt M},
\end{align}
where $\kappa\approx 4.9$.
The points $n_-$ and $n_+$ are those between which the state coefficients rapidly increase. There is also a second region for large $n$ where the state coefficients rapidly decrease. The symmetry of the distribution implies that the sum over that region has the same lower bound. Using Eq.~(\ref{var_from_psi}), and the fact that $V_{\rm H}\ge V_{\rm C}$, the variance for sufficiently large $N$ is lower bounded by
\begin{equation}
V_{\rm H}(\bestest)\ge \frac 1{\kappa(N_K+1)\sqrt M}.
\end{equation}
The smallest possible value for this lower bound is obtained for both $N_K$ and $M$ of order $N$, which gives
\begin{equation}
V_{\rm H}(\bestest) = \Omega(N^{-3/2}).
\end{equation}
\end{proof}

In Ref.\ \cite{Higgins08} it was proven that measurements could yield a variance scaling as $O(\sqrt{\ln N}N^{-3/2})$. The lower-bound scaling here is the same, except for a small $\sqrt{\ln N}$ factor. The method used for the proof in \cite{Higgins08} was to consider an analysis of the data where, if the phase estimates from the single-photon measurements and the QPEA differed by too large an amount, the phase estimate from the single-photon measurements was used, and otherwise the phase estimate from the QPEA was used.

The general idea behind the proof of \cite{Higgins08} is that the phase estimate from single-photon measurements has a negligible probability of having error larger than $O(1/\sqrt M)$. Therefore, if the phase estimates from the single-photon measurements and the QPEA differ by more than $O(1/\sqrt M)$, it is almost certainly the QPEA phase estimate that is wrong, so the phase estimate from the single-photon measurements should be used. The variance then is of order $1/M$, but because it occurs with probability $\sqrt{M}/N_K$, the total contribution to the variance is $O(1/(N_K\sqrt{M}))$. If the phase estimates do agree to within $O(1/M)$, then the contribution to the variance is again of order $1/(N_K\sqrt{M})$.

The additional factor of $\sqrt{\ln N}$ comes about because it is not quite true that the probability of the single-photon measurements having error larger than $O(1/\sqrt M)$ can be ignored. This probability must scale down with $M$ to prevent it giving a contribution to the variance larger than $O(1/(N_K\sqrt{M}))$. To ensure that this probability is sufficiently small, the size of the error should be $O(\sqrt{(\ln M)/M})$. This leads to the additional $\sqrt{\ln N}$ factor in the final result.

\begin{figure}
\centering
\includegraphics[width=0.48\textwidth]{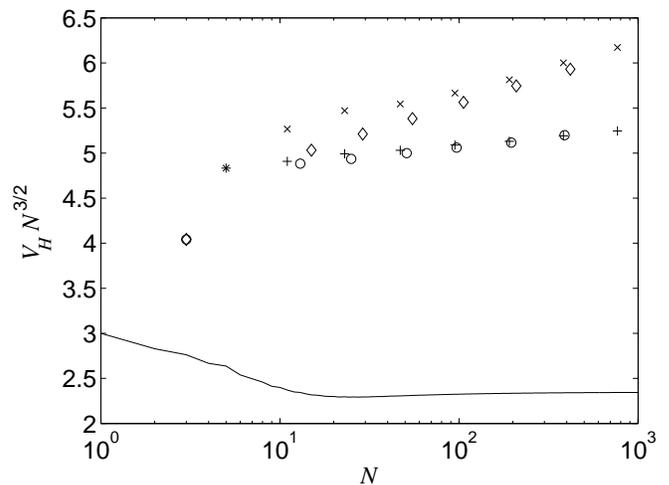}
\caption{The variance for hybrid measurements. The solid line is the canonical phase variance for the equivalent two-mode state. The crosses and pluses are for $M=2^K$, and increments in $\feedphase$ of $\pi/M$ and $\pi/2$, respectively. The diamonds and circles are for values of $M$ that give the minimum value of $V_{\rm H}N^2$, and increments in $\feedphase$ of $\pi/M$ and $\pi/2$, respectively.}
\label{fig:hybrid}
\end{figure}

The analytical results show that there is a lower bound on the variance for hybrid measurements of $\Omega(N^{-3/2})$, and an upper bound of $O(\sqrt{\ln N}N^{-3/2})$. The scaling must be close to $N^{-3/2}$, but these bounds leave open the question of whether there is an additional logarithmic factor. The hybrid measurements have been simulated for values of $K$ up to 8, and the results are shown in Fig.\ \ref{fig:hybrid}. The variances shown are multiplied by $N^{3/2}$ to make the scaling clearer. For the results shown the value of $V_{\rm H}N^{3/2}$ increases with $N$, suggesting that the variance is slightly larger than $O(N^{-3/2})$.

In Fig.\ \ref{fig:hybrid}, results are shown for increments in $\feedphase$ of $\pi/M$ and $\pi/2$. The variance for the single-photon measurements on their own is slightly smaller with the $\pi/M$ increments than with the $\pi/2$ increments. However, for the hybrid measurements, using increments of $\pi/2$ yields significantly better results. Results are also shown for $M=2^K$, as well as for where the value of $M$ has been adjusted to minimize $V_{\rm H}N^{3/2}$. The motivation for using $M=2^K$ is that the analytical results suggest that $M\approx N/3$ is optimal. Numerically it was found that slightly higher values of $M$ gave slightly better results, particularly for $\feedphase$ increments of $\pi/M$.

\subsection{Nonadaptive scheme} \label{sec:nonadapt}
It might be expected that approaching the Heisenberg limit requires an adaptive measurement. Normally the variance would be approximately the sum of the canonical phase variance and an additional variance due to the measurement technique \cite{Wiseman97}, and nonadaptive measurements would introduce a variance scaling as the SQL \cite{twomodepra}. Despite this, it is still possible to achieve the same scaling as the Heisenberg limit with a nonadaptive technique \cite{Higgins08}. The reason why this is possible is that the nonadaptive measurement with multiple time modes has a different POVM from the nonadaptive measurement on a single time mode.

A new feature of the nonadaptive technique of \cite{Higgins08} is that the number of repetitions, $M$, is now a function of the values of $K$ and $k$. The largest value of $M$ is for a single-photon NOON state, and the number of repetitions is decreased as the size of the NOON state (or number of passes in Ref.\ \cite{Higgins08}) is increased. The reason for changing the number of repetitions is essentially that the role of the measurements for smaller values of $k$ is to distinguish between the multiple phase estimates provided by larger values of $k$. If the measurements for smaller values of $k$ do not distinguish between these phase estimates accurately, then the resulting error is large. On the other hand, because the resource cost for the smaller values of $k$ is small, the cost of repeating these measurements is low. It is therefore better to repeat these measurements more often, in order to prevent these large errors.

\begin{figure}
\centering
\includegraphics[width=0.48\textwidth]{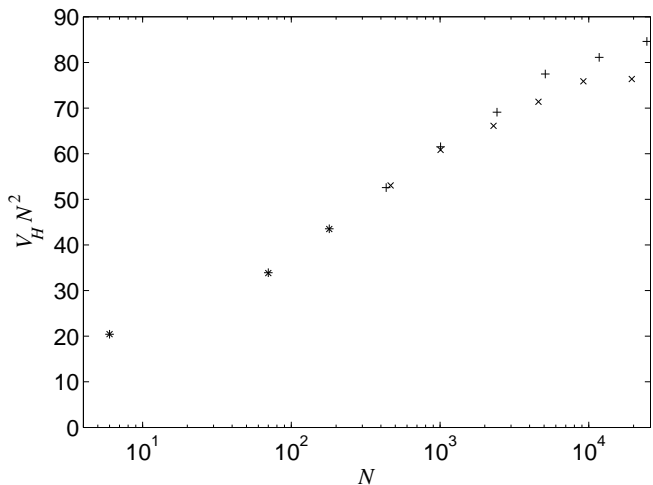}
\caption{The variance for nonadaptive measurements with $M$ optimized as a function of $K$, but independent of $k$. The crosses are for $2^{20}$ samples and the pluses are for $2^{25}$ samples.}
\label{fig:nonad}
\end{figure}

If the number of repetitions is not changed with $k$, then the numerical results indicate that the scaling of the variance is not as good as $1/N^2$. The numerical results are shown in Fig.\ \ref{fig:nonad}. The value of $V_{\rm H}N^2$ increases approximately linearly in $\ln N$, indicating that the scaling is as $(\ln N)/N^2$. For these results the value of $M$ was increased with $K$, even though it was independent of $k$. It was found that the best value of $M$ increased nearly linearly in $K$, from about 10 for $K=2$ to 24 for $K=9$. The primary contribution to the variance was from low-probability results with large error. This meant that extremely large numbers of samples need to be used to obtain accurate estimates of the variance. Calculations with $2^{20}$ samples yielded lower estimates of the variance than those for $2^{25}$ samples, because they did not sufficiently sample the low probability results.

The reason why $M$ needs to be increased with $K$ in the nonadaptive case can be understood in the following way. For simplicity, consider just the measurements with $k=0$ (a single-photon NOON state), and assume that the measurements with larger values of $k$ are sufficient to narrow down the estimates of the phase to $\sysphase$ and $\sysphase+\pi$. The measurements with $k>0$ measure the phase modulo $\pi$, so they cannot distinguish between these alternatives. They do not give the phase exactly modulo $\pi$, but assuming that they do can only decrease the phase variance. The measurements with $k=0$ could distinguish between the two alternatives if they were known.

In the adaptive case, the phase is known accurately (though not perfectly) modulo $\pi$ at the stage the measurements with $k=0$ are performed. The feedback phase $\feedphase$ can therefore be chosen so as to efficiently distinguish between the two alternatives. However, in the \emph{nonadaptive} case, the values of $\feedphase$ used are completely independent of the actual phase. In the case where there is a fixed value of $M$, there is a nonzero probability that the measurement results will be more consistent with $\sysphase+\pi$ rather than $\sysphase$, so there is a $\pi$ error in the phase estimate.

In particular, the probability that all the measurement results with $k=0$ are consistent with $\sysphase+\pi$ rather than $\sysphase$ is
\begin{equation}
P_{\rm error} = \prod_{m=1}^M \frac 12 (1-|\cos(\sysphase-\feedphase_m)|).
\end{equation}
There will be other combinations of measurement results that are also more consistent with $\sysphase+\pi$, but the probability for this combination on its own provides a lower bound on the overall probability that the final phase estimate is $\sysphase+\pi$. For some particular values of the system phase this probability will be zero, but averaging over all system phases gives a nonzero error probability.

This lower bound on the probability of a measurement result that gives a $\pi$ error in the phase estimate means that if nonadaptive measurements are performed with a fixed value of $M$, the phase variance can not decrease below some minimum value. More generally, if the value of $M$ is allowed to depend on $K$ (but still be independent of $k$), then there need not be a fixed lower bound on the variance, but because of Eq.\ \eqref{eq:repsbnd} the variance can scale no better than $M(N)/N^2$. As $M$ must increase with $K$ (and therefore $N$) in order to remove the lower bound on the variance, the variance does not scale at the Heisenberg limit.

\begin{figure}
\centering
\includegraphics[width=0.48\textwidth]{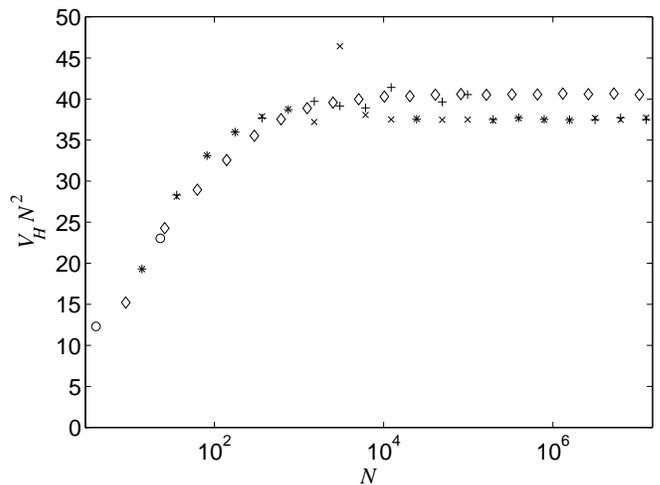}
\caption{The variance for nonadaptive measurements with $M$ a function of both $K$ and $k$. The results for $M(K,k)=2+3(K-k)$ are shown as diamonds, the results for $M(K,k)=4+2(K-k)$ are shown as both the pluses and crosses. The crosses are for $2^{20}$ samples and the pluses are for $2^{25}$ samples. The circles are $M(K,k)=1+K-k$ for $K=1$ and $M(K,k)=1+4(K-k)$ for $K=2$.}
\label{fig:nonadmu}
\end{figure}

For the nonadaptive scheme given in Ref.\ \cite{Higgins08} the variation of $M$ used was $M(K,k)=2+3(K-k)$. Numerically it was found that this variation of $M$ gave the smallest asymptotic variance that could be reliably calculated. The value of $V_{\rm H}N^2$ was approximately $40.5$ for values of $N$ up to more than $10^7$ (see Fig.\ \ref{fig:nonadmu}). This corresponds to an uncertainty approximately $2.03$ times the Heisenberg limit.

Alternative values of $M$ given by $M(K,k)=4+2(K-k)$ did give slightly smaller values of $V_{\rm H}N^2$ for large $N$. However, calculations in this case were not reliable because the variance depends heavily on results with large error and small probability. Using $2^{20}$ samples gave significantly smaller values of $V_{\rm H}N^2$ in most cases, except for one value which was much larger. When the number of samples was increased to $2^{25}$ the value of $V_{\rm H}N^2$ increased to approximately the same as for $M(K,k)=2+3(K-k)$ for $N$ up to about $10^5$. For larger values of $N$ the variance was still smaller, but it is likely that the number of samples is still too small to obtain a reliable result.

Using other functions for $M(K,k)$ did give slightly better results for small values of $K$. For $K=1$, the functional dependence that gave the smallest value of $V_{\rm H}N^2$ was $M(K,k)=1+K-k$. For $K=2$, the best values of $M$ were given by $M(K,k)=1+4(K-k)$. The results for these cases are also shown in Fig.\ \ref{fig:nonadmu}.

\section{Adapting the size of the NOON state}
\label{sec:size}
Up to this point we have considered a fixed sequence of NOON states, and the only parameter which was varied adaptively was the feedback phase. It is also possible to consider adjusting the size of the NOON states used according to the measurement results. The most obvious way of doing this is to select the NOON state that minimizes the expected variance after the next detection. The problem with this approach is that it will select large NOON states that give small variance, but result in a greater use of resources. As the aim is to obtain the smallest possible scaling constant for $V_{\rm H}N^2$, a better approach is to choose the NOON state that minimizes this quantity.

\begin{figure}
\centering
\includegraphics[width=0.48\textwidth]{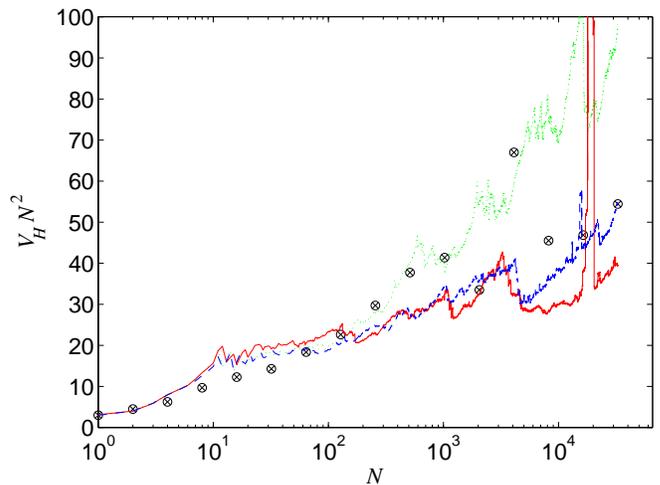}
\caption{The variance for measurements where the size of the NOON states are adapted. The crossed circles are where the size is chosen to minimize the expected value of $V_{\rm H}N^2$ after the next detection, and the sizes are chosen such that a particular value of $N$ is obtained. The continuous lines are where all intermediate values of $N$ are sampled. The dotted green line is where $V_{\rm H}N^2$ is minimized, the solid red line is where $[\lr S -\ln (2\pi)]/ \ln N$ is minimized, and the dashed blue line is where $V_{\rm H}N^2$ is minimized, but the feedback of \cite{Mitchell2005} is used for the phase (color online).}
\label{fig:adn}
\end{figure}

To determine results for a specific value of $N$, the sizes of the NOON states were restricted such that the total resources did not exceed $N$. The resulting variances obtained are shown as the crossed circles in Fig.\ \ref{fig:adn}. For the larger values of $N$ ($2^{11}=2048$ and above), it was only feasible to use $2^{10}$ samples, so these results have low accuracy. Nevertheless it is clear that using this approach yields a significantly higher variance than simply using adaptive measurements with a fixed sequence of NOON states. In addition, the scaling of the variance appears to have increased to $(\ln N)/N^2$.

It is possible to improve on this adaptive scheme by using a radically different approach \cite{Mitchell2005}. There are three main features of this approach:
\begin{enumerate}
\item The phase estimation begins with 100 measurements with single photons and no feedback 
(similar to the hybrid technique introduced in Sec.~\ref{sec:Hybrid}).
\item After the first step, the feedback phase is chosen such that the expected probabilities for the two measurement results are equal.
\item After the first step, the size of the NOON state is chosen to minimize the expected entropy of the phase distribution after the next measurement.
\end{enumerate}
More specifically, the size of the NOON states is chosen to minimize $\lr S / \ln N$, where $N$ is the number of resources after the next detection. The motivation for dividing by $\ln N$ can be understood in the following way. For a probability distribution, the negative of the entropy approximately corresponds to the information that may be communicated using a distribution of that form. As the distribution narrows by a factor of two, the phase is effectively known to an additional bit, and the entropy decreases by 1. Therefore the entropy is approximately proportional to the log of the phase uncertainty. The phase uncertainty varies as a power of $N$, so the entropy should be approximately proportional to the log of $N$. The best phase information for a given amount of resources can therefore be obtained by minimizing $\lr S / \ln N$.

The calculations given in Ref.\ \cite{Mitchell2005} also used some additional simplifications to increase the speed. Rather than searching all possible values of the size of the next NOON state (which could be many thousands), only about 30 were searched. Denoting the number of resources used so far by $N$, only sizes up to $\lceil N/3 \rceil$ were searched, and in step sizes of $\lceil N/100 \rceil$. In addition, each run was used to obtain phase estimates for each intermediate number of resources. There is a complication in doing this, because many of the NOON states have $\nm>1$, so there are not phase estimates for all values of $N$. To obtain phase estimates for all $N$, the estimates 
for these skipped values were approximated by taking the phase estimate from the next lower value of $N$ that was not skipped. This gives samples for all intermediate values of $N$, and cannot underestimate the variance. Even with these simplifications, the phase variance scaled approximately as $40/N^2$.

On further investigation, we have found that the feedback scheme of Ref.\ \cite{Mitchell2005} can be further simplified, and provide more accurate phase estimates. One simplification is in the approximation used for the entropy. It is not practical to calculate the entropy exactly, but it can be accurately approximated from the moments of the distribution. Ref.\ \cite{Mitchell2005} used a higher-order approximation involving the third- and fourth-order moments (the approximation given in Eq.\ (10) of \cite{Mitchell2005} is for the negative of the entropy). The calculation can be simplified by using the simpler approximation
\begin{equation}
\label{eq:simpler}
S \approx \frac 12 [1+\ln(2\pi) + \ln V_{\rm C}],
\end{equation}
Numerically it is found that this simpler approximation gives accuracy of the phase measurements similar to that using the higher-order approximation.

It was also found that the variance could be reduced if the quantity that was minimized was of the form $[\lr S + C]/ \ln N$, where $C$ is some constant. One value that gave particularly good results was $C = -\ln (2\pi)$ (which just corresponds to changing the sign on $\ln(2\pi)$ in Eq.\ \eqref{eq:simpler} before dividing by $\ln N$). In Ref.\ \cite{Mitchell2005} the measurement scheme was started with 100 single-photon measurements, in order to obtain a narrowly peaked distribution. The drawback to this approach is that it increases the variance for total photon numbers of a similar order. The majority of the resources need to be used in NOON states with high photon numbers in order to obtain results near the Heisenberg limit. To reduce this problem, the number of single-photon measurements was reduced to 10.

In Fig.\ \ref{fig:adn} the results from three different approaches are shown. The first scheme described in this section, except with the sampling scheme of Ref.\ \cite{Mitchell2005}, is shown as the dotted line. This sampling appears to give slightly worse results in this case as compared to the results where a fixed maximum value was used. For larger values of $N$, $V_{\rm H}N^2$ is typically above 60. The results where $[\lr S -\ln (2\pi)]/ \ln N$ was minimized are also shown in this figure. This seemed to give the best results out of the alternative variations of the scheme of Ref.\ \cite{Mitchell2005} that were tested, with variances significantly below those where $V_{\rm H}N^2$ was minimized.

The third alternative that is shown in Fig.\ \ref{fig:adn} is that where $V_{\rm H}N^2$ is minimized when choosing the size of the NOON state, but the feedback phase is chosen to ensure that the probabilities of the two detection results are equal (as in Ref.\ \cite{Mitchell2005}). This can be achieved simply by replacing $[\lr S +C]/ \ln N$ with $\lr S + 2\ln N$. This scheme also gives better results than the first scheme, and gives variances close to those for the scheme minimizing $[\lr S -\ln (2\pi)]/ \ln N$.

Even though the schemes where the size of the NOON states are adapted are far more flexible than the schemes with a fixed sequence of NOON states, they do not appear to give any better results. For most methods tested the variance obtained is far greater than that for a fixed sequence of NOON states. Even for the best method tested, the variances obtained were simply comparable to those for the fixed sequence of NOON states, and did not give any improvement. Because the fixed sequence of NOON states is far simpler to implement, it would be preferable to use for most applications.

\section{Conclusions}
\label{sec:conc}
NOON states achieve the highest phase resolution possible for a given photon number, but do not provide an unambiguous phase estimate. By combining measurements with a range of NOON states with different photon numbers, it is possible to achieve a phase resolution at the Heisenberg limit, and eliminate the ambiguity in the phase estimate. As well as being applicable to NOON states, these schemes can also be performed using single photons, by passing them through the phase shift multiple times, as experimentally demonstrated in Refs.\ \cite{Higgins07,Higgins08}.

In this paper we have considered many different measurement schemes involving sequences of NOON states, some of which give a variance 
scaling at the Heisenberg limit, and some of which do not. Below, we discuss these results, but for the benefit of the reader we present a summary 
of all the key results in this paper in Table~\ref{sumtab}. 

\begin{table*}[htdp] 
\begin{tabular}{ |c|c|c|r|c|l|r| }
\hline Scheme  & NOON Sequence & Intrinsic Var.  &  Reference\;\;\ & Estimate Var.  & Rating & Reference\;\;\; \\
\hline single photons  & $1,1, \cdots, 1$ & $\Or(1/N)$  & & $\Or(1/N)$ &*&   \\ 
\hline  QPEA:  $M=1$ & $2^K,2^{K-1}, \cdots, 1$ & $\Or(1/N)$ & \cite{Higgins07}; Sec.~IV  & $\Or(1/N)$  & * & \cite{Higgins07}; Sec.~V \\
\hline generalized QPEA $M=2$ & $2^K$ twice, $ \cdots, 1$ twice & $\Or(\ln N / N^2)$  & Sec.~IV & $\Or(1/N)$  & * &\cite{Higgins07}; Sec.~V  \\
\hline generalized QPEA $M=3$ & $2^K$ thrice, $ \cdots, 1$ thrice & $\Or(1/N^2)$ & Sec.~IV   & $\Or^*(1/N^{3/2})$  &*** &  Sec.~V   \\
\hline generalized QPEA $M\geq 4$ & $2^K$ $M$-fold, $ \cdots, 1$ $M$-fold & $\Or(1/N^2)$ & Sec.~IV   & $\Or^*(1/N^{2})$  &***** &  \cite{Higgins07}; Sec.~V   \\
\hline generalized QPEA $K$ fixed & $2^K$ $M$-fold, $ \cdots, 1$ $M$-fold & $\Or(2^{-2K}/N)$ & Sec.~VI   & $\Or^*(2^{-2K}/N)$  &  ** & Sec.~VI   \\
\hline nonadaptive, best $M(K)$  & $2^K, \cdots, 1$, each $M(K)$-fold & $\Or(M(K)/N^2)$ & Sec.~VII   & $\Or^*(\ln N / N^2)$  &  **** & Sec.~VII   \\
\hline nonadaptive, best $M(K,k)$ & $\{2^k\}$, each $M(K,k)$-fold & $\Or(1/N^2)$ & \cite{Higgins08}; Sec.~VII   & $\Or(1/N^{2})$  &   ***** &\cite{Higgins08}; Sec.~VII   \\
\hline  hybrid QPEA+singles & $2^K,2^{K-1}, \cdots, 1,1,\cdots, 1$ & $\Omega(1/N^{3/2})$ & Sec.~VII   & $O(\sqrt{\ln N}/N^{3/2})$  & *** & \cite{Higgins08}; Sec.~VII   \\
\hline  non-binary adaptive  & adaptively chosen & N/A &   & $\tilde{O}^*(1/N^{2})$ & **** & \cite{Mitchell2005}; Sec.~VIII   \\
\hline \end{tabular} 
\caption{Scalings of the phase variance for the various phase estimation schemes considered in this paper. The column ``NOON Sequence'' shows the number of photons in each state in the sequence, in temporal order (which is important for adaptive measurements). In all but the last scheme, these photon numbers are powers of $2$ between $1$ (a single-photon state) and $2^K$. Two variances are shown: the intrinsic variance (i.e., for a canonical phase measurement), 
and the variance for the actual measurement scheme. The former is undefined for the final scheme because the sequence of states there is not predetermined. Rather than exact variances, only the scaling of the variance with $N$, the total number of photons, is shown. The $\Or$, $O$ and $\Omega$ are standard Bachmann-Landau notation; $\Or$ means the asymptotic scaling, while $O$ and $\Omega$ mean an upper and lower bound, respectively, on the asymptotic scaling. The symbol $\tilde{O}$ indicates that multiplying factors scaling as powers of $\log(N)$ are ignored. We use an asterisk, such as in $\Or^*$, to indicate that the scaling is only proven numerically. For each result we give the section where the result is discussed, as well as a citation to previous work (if it exists). Finally, the ``Rating'' column gives an easy guide to the performance of the various phase estimation schemes. The stars have the following interpretation:
* --- variance equal to, or a constant multiplier greater than, the SQL of $1/N$ asymptotically;
** --- variance scaling as the SQL, but smaller by a constant multiplier;
*** --- variance scaling roughly as the geometric mean of the SQL, and the Heisenberg-limit (HL) of $\pi^2/N^2$ asymptotically;
**** --- variance scaling almost as well as the HL;
***** --- variance equal to, or a constant multiplier times greater than, the HL. 
  \label{sumtab}}
\end{table*}

Using a sequence of NOON states with photon numbers that are decreasing powers of two, an elegant adaptive optical phase measurement can be performed by implementing the quantum phase estimation algorithm from quantum computing theory \cite{Cleve1998,Nielsen2000}. This provides a canonical measurement of the phase, which is a remarkable result because normally it is impossible to perform a canonical phase measurement using linear optics. The canonical measurement is possible because the photons are separated into distinct time modes.

This approach does not provide variance scaling as the Heisenberg limit, because the underlying state has poor phase properties, and gives a significant probability for large phase errors. Reduced phase variance could be achieved using states that are entangled between the different time modes. A more practical alternative is to use multiple copies of the state, which also provides reduced canonical phase variance. The canonical measurements can no longer be performed, but for 4 or more copies it is possible to accurately approximate the canonical phase measurement using adaptive measurements. The minimum variance is then obtained using 5 copies.

Perhaps surprisingly, it is also possible to achieve the Heisenberg limit using nonadaptive measurements. If the same number of copies of each size of NOON state is used, then the phase uncertainty scales as $\sqrt{\ln N}/N$. In order to achieve the Heisenberg limit, the number of copies must be a function of the size of the NOON state.
In particular, the number of copies must be of order unity (e.g. 2) for the largest NOON state, and should increase by a constant number (e.g. 3) at each step as one halves the size of the NOON state.  

An alternative to fixing the number of copies of a given NOON state, is to consider schemes which choose the size of the next NOON state to be used adaptively. This approach can yield a variance close to the Heisenberg limit.
The obvious approach, which is to choose a NOON state which minimizes the variance multiplied by $N^2$, provides scaling slightly worse than the Heisenberg limit. However, it is possible to obtain scaling much closer to the Heisenberg limit using an alternative scheme where the entropy of the phase distribution is minimized instead. This approach was first presented in Ref.\ \cite{Mitchell2005}. Here we have simplified and improved that scheme in a number of ways. 

Some other alternatives that might be expected to provide phase measurements at the Heisenberg limit do not. This can be conveniently proven by examining canonical measurements for the equivalent two-mode state. One such alternative is hybrid measurements, where the QPEA is supplemented with single-photon detections, to reduce the probability of large phase errors. This approach yields a canonical phase uncertainty scaling as $1/N^{3/4}$, and the uncertainty in the phase estimate from the hybrid measurements scales slightly worse. Another alternative that does not yield the Heisenberg limit is simply increasing the number of repetitions, rather than the maximum size of the NOON state (or, equivalently, number of passes). While this gives results which remain below the standard quantum limit of $1/\sqrt{N}$ for phase precision, they {\em scale} in exactly the same way as the SQL.

In this paper we have demonstrated this latter result experimentally by keeping the maximum number of passes fixed while increasing the number of photons used. This demonstration helps illustrate why gravitational wave interferometers do not approach the limit of $1/N$ for phase precision even though they use multiple passes: in such interferometers the number of passes  is {\em fixed} at its maximum value (which is typically very large), while the number of ``repetitions'' (the number of photons in the coherent state which is used) is also made very large. This contrasts with all of the schemes that do scale as $1/N$, in which the number of repetitions is of order unity for the largest pass-number (or largest NOON state), and in which the number of passes (or size of the NOON state) must be varied across a large range of values from one to its maximum value.

\acknowledgments
This work was supported by the Australian Research Council and Spanish MEC projects QOIT (Consolider-Ingenio 2010) and FLUCMEM (MWM). We would also like to thank R. Mu\~{n}oz-Tapia, K. Resch and M. Mosca for helpful discussions.

\appendix

\section{Proof of Theorem \ref{th:mult}}
\label{app}

Here we give the full proof of Theorem \ref{th:mult}, which gives the scaling of the canonical phase variance for repetitions of the measurements. In general the $f_M(n)$ can be obtained by the recurrence relation
\begin{equation}
f_{M}(n) = \sum_{k=0}^{N_K} f_{M-1}(n-k).
\end{equation}
Note that this is a discrete convolution of $f_{M-1}$ with the uniform distribution. This recurrence relation comes about because the number of combinations of the values of $n_1,n_2,\ldots,n_M$ may be obtained, for each value of $n_1$, by determining the number of combinations of $n_2,\ldots,n_M$ that sum to $n-n_1$. The simplest example is $M=1$, where $f_1(n)=1$ for $0\le n\le N_K$ and $f_1(n)=0$ otherwise. For $M=2$, evaluating the sum yields
\begin{equation}
f_2 (n) = \left\{ {\begin{array}{*{20}l}
   {n + 1,} & {0 \le n \le N_K}  \\
   {2N_K - n + 1,} & {N_K \le n \le 2N_K}  \\
   {0,} & {\rm otherwise.}  \\
\end{array}} \right.
\end{equation}
A crucial point is that the increment between successive values of $f_M$ is bounded. The recurrence relation implies
\begin{align}
&|f_{M}(n+1)-f_{M}(n)| \nn & \quad \le (N_K+1)\max_{k} |f_{M-1}(k+1)-f_{M-1}(k)|.
\end{align}
Because $|f_2(n+1)-f_2(n)|\le 1$, this means that
\begin{equation}
|f_{M}(n+1)-f_{M}(n)| \le (N_K+1)^{M-2}.
\end{equation}
The function is symmetric, so $f_M(n)=f_M(N-n)$. It is possible to determine the general value of $f_M(n)$ for $n\le N_K$. The solution is
\begin{align}
f_M(n) = {{n+M-1}\choose{M-1}} = \frac{(n+M-1)!}{n!(M-1)!}.
\end{align}
To prove this, first note that it is correct for $M=2$. Next, use the recurrence relation to give
\begin{align}
&f_{M}(n)-f_{M}(n-1) \nn &= \sum_{k=0}^{N_K} f_{M-1}(n-k)- \sum_{k=0}^{N_K} f_{M-1}(n-k-1) \nn
&= \sum_{k=0}^{N_K} f_{M-1}(n-k)- \sum_{k=1}^{N_K+1} f_{M-1}(n-k) \nn
&= f_{M-1}(n)-f_{M-1}(n-N_K-1).
\end{align}
For $n\le N_K$, the argument of $f_{M}(n-N_K-1)$ is negative, so this term is zero. If the solution is correct for $M-1$ and for $n-1$, then
\begin{align}
f_{M}(n) &= f_{M}(n-1)+f_{M-1}(n) \nn
&= \frac{(n+M-2)!}{(n-1)!(M-1)!} + \frac{(n+M-2)!}{n!(M-2)!} \nn
&= (n+M-1)\frac{(n+M-2)!}{n!(M-1)!} \nn
&= \frac{(n+M-1)!}{n!(M-1)!}.
\end{align}
Using the formula gives, for $n<N_K$,
\begin{align}
f_M(n+1)-f_M(n) & \le (n+2)^{M-2}, \\
f_M(n+1)+f_M(n) & \ge \frac{(n+2)^{M-1}}{(M-1)!}.
\end{align}
For $M>2$, this gives
\begin{equation}
\frac{[f_M(n+1)-f_M(n)]^2}{[f_M(n+1)+f_M(n)]} \le (M-1)!(N_K+1)^{M-3}.
\end{equation}
Due to symmetry, the same result holds for $n>N-N_K$. This part of the derivation also holds for $n=-1$ or $N$. For $M=2$ or 1, a decreasing function of $n$ is obtained, so the $n$ cannot be replaced with $N_K$ for the upper bound.

The distribution for $f_M(n)$ cannot have a nontrivial minimum (the trivial minimum is zero at the bounds). That is, it cannot decrease with $n$, then increase. This can be shown by induction in the following way. If and only if $f_M(n)$ has a nontrivial minimum, then there exist $n$, $n'$, $\delta n>0$ and $\delta n'>0$ such that $n'<n$, $f_{M}(n)>f_{M}(n-\delta n)$ and $f_{M}(n')<f_{M}(n'-\delta n')$. That would imply $f_{M-1}(n)>f_{M-1}(n-N_K-\delta n)$ and $f_{M-1}(n')<f_{M-1}(n'-N_K-\delta n')$. However, that would imply that $f_{M-1}(n)$ has a nontrivial minimum. As we know that is not the case for $M=1$ and 2, using induction shows that $f_M(n)$ cannot have a nontrivial minimum for any $M$.

This result implies that $f_M(n)$ increases to a maximum at the centre, then decreases. In particular, for $N_K\le n\le N-N_K$, $f_M(n)\ge f_M(N_K)$. Now we have $f_M(N_K)\ge (N_K+1)^{M-1}/(M-1)!$, so
\begin{equation}
\frac{[f_M(n+1)-f_M(n)]^2}{[f_M(n+1)+f_M(n)]} \le (M-1)!(N_K+1)^{M-3},
\end{equation}
for $N_K< n\le N-N_K$. Therefore, overall
\begin{equation}
\sum_{n=-1}^{N_K} \frac{[f_M(n+1)-f_M(n)]^2}{[f_M(n+1)+f_M(n)]} \le M!(N_K+1)^{M-2}.
\end{equation}

The variance $V_{\rm C}$ for the state is given by
\begin{align}
&2(1-|\lr{e^{i\bestest}}|) \nn &= \frac{1}{(N_K+1)^M}\sum_{n=-1}^{N}\left(\sqrt{f_M(n+1)}-\sqrt{f_M(n)}\right)^2 \nn
&\le \frac{1}{(N_K+1)^M}\sum_{n=-1}^{N}\frac{(f_M(n+1)-f_M(n))^2}{(f_M(n+1)+f_M(n))} \nn
&\le \frac{M!}{(N_K+1)^2} = \frac{M!M^2}{(N+M)^2}.
\end{align}
Hence for $M>2$, the variance scales as the Heisenberg limit of $1/N^2$.

In the case $M=2$, the bound for $n<N_K$ is
\begin{equation}
\frac{
[f_M(n+1)-f_M(n)]^2}{[f_M(n+1)+f_M(n)]} \le \frac 1{n+1}.
\end{equation}
Taking the sum from $n=0$ to $N_K-1$ gives
\begin{equation}
\sum_{n=0}^{N} \frac{[f_M(n+1)-f_M(n)]^2}{[f_M(n+1)+f_M(n)]} \le \int_{0}^{N_K} \frac {dn}{n+1}
= \ln(N_K+1).
\end{equation}
This approach therefore yields an upper bound on the variance that scales as $(\ln N)/N^2$, rather than the Heisenberg limit of $1/N^2$. In fact, this is the actual scaling for these states. Evaluating the exact value of $V_{\rm C}$ gives
\begin{align}
2(1-|\lr{e^{i\bestest}}|) &= 2\left(1-\frac{2}{(N_K+1)^2}\sum_{n=1}^{N_K} \sqrt{n(n+1)} \right) \nn
&= \frac {2\ln N}{N^2} + \frac {c}{N^2} + O(N^{-3}).
\end{align}
where $c\approx 6.5949$.

\section{Details of sharpness formula}
\label{sec:det}
Here we give further details of the meaning of Eq.\ \eqref{eq:remu} and its interpretation as an average over the sharpnesses of the individual Bayesian probability distributions obtained. A subtlety that has been omitted in the discussion in the main text is that the probability distribution for the measurement results is now a function of both the system phase and initial feedback phase, and should therefore be written $P(\vec u_m|\sysphase,\feedphase_1)$. The other feedback phases are all chosen deterministically based on the measurement results and $\feedphase_1$, and can be omitted here. We use the notation $\feedphase_j$ to indicate the feedback phase before the $j$'th detection. The value of $\mu$ may then be written explicitly as
\beq
\mu =  \left| \frac 1{2\pi}\int d\feedphase_1 \sum_{\vec u_m} P(\vec u_m|\sysphase,\feedphase_1) e^{i\bestest}\right|.
\eeq

Because the phases are relative, the probability of the measurement results depends only on the difference between the system phase and the first feedback phase. That is,
\begin{equation}
P(\vec u_m|\sysphase,\feedphase_1)=P(\vec u_m|\sysphase-\feedphase_1,0).
\end{equation}
Similarly the probability distribution for the system phase will depend on $\feedphase_1$, but will depend only on the difference between the two; that is,
\begin{equation}
P(\sysphase|\vec u_m,\feedphase_1)=P(\sysphase-\feedphase_1|\vec u_m,0).
\end{equation}
In the remainder of this appendix we use the probabilities in this form, and omit the argument of $0$.

The later feedback phases, as well as the phase estimate $\bestest$, depend on the measurement results and the value of $\feedphase_1$. Because the phases are relative, these phases should only depend in a linear way on the initial feedback phase. That is,
\beq
\bestest(\vec u_m,\feedphase_1) = \bestest(\vec u_m,0)+\feedphase_1.
\eeq
Using this, and changing variables to $\dphi = \sysphase-\feedphase_1$, we find that $\mu$ becomes
\beq
\mu =  \left| \frac 1{2\pi}\int d\dphi \sum_{\vec u_m} P(\vec u_m|\dphi,0) e^{i(\bestest(\vec u_m,0)-\dphi)}\right|.
\eeq

In the following analysis we need to consider a set of measurement results given an actual system phase $\sysphase$ and initial feedback phase $\feedphase_1$, as well as the Bayesian probability distribution for the system phase given those measurement results. Since we effectively have two system phases, the actual system phase and the dummy variable for the system phase in the Bayesian probability distribution, we use the symbol $\sysphase_{\rm B}$ for the variable in the Bayesian probability distribution. We also use $\dphi_{\rm B}\equiv\sysphase_{\rm B}-\feedphase_1$.
Then the optimal value of $\bestest$ is
\begin{align}
\bestest(\vec u_m,\feedphase_1) &= \feedphase_1 + \arg \int e^{i\dphi_{\rm B}} P(\dphi_{\rm B}|\vec u_m) d\dphi_{\rm B} \nonumber \\
&= \arg \int e^{i\dphi_{\rm B}+\feedphase_1} P(\dphi_{\rm B}|\vec u_m) d\dphi_{\rm B} \nonumber \\
&= \arg \int e^{i\sysphase_{\rm B}} P(\sysphase_{\rm B}-\feedphase_1|\vec u_m) d\sysphase_{\rm B} \nonumber \\
&= \arg \lr {e^{i\sysphase_{\rm B}}},
\end{align}
where the angle brackets indicate that the average is taken for that value of $\vec u_m$ and $\feedphase_1$.
The integral for $\mu$ then becomes
\begin{equation}
\mu = \frac 1{2\pi}\sum_{\vec u_m} \left|\int e^{i\dphi}P(\vec u_m|\dphi)d\dphi \right|.
\end{equation}
In this expression the integral over $\dphi$ can be an integral over the actual system phase for fixed $\feedphase_1$, or an integral over the feedback phase $\feedphase_1$ for fixed $\sysphase$. For Eq.\ \eqref{eq:remu} in the main text, we have taken $\feedphase_1=0$ so $\dphi=\sysphase$.

This expression can be rewritten as
\begin{equation}
\mu =\frac 1{2\pi}\int d\feedphase_1 \sum_{\vec u_m} P(\vec u_m|\sysphase_{\rm B}-\feedphase_1) \left|\int e^{i\dphi}P(\dphi|\vec u_m)d\dphi \right|.
\end{equation}
The interpretation of the variables in this expression now changes. We now interpret this expression as the sharpness of the Bayesian probability distribution for the system phase given measurement results $\vec u_m$ averaged over the initial feedback phase $\feedphase_1$ and measurement results $\vec u_m$, for a specific system phase. To make this more clear, we can relabel the variables as below
\begin{align}
\mu &=\frac 1{2\pi}\int d\dphi \sum_{\vec u_m} p(\vec u_m|\dphi) \left|\int e^{i\dphi_{\rm B}}P(\dphi_{\rm B}|\vec u_m)d\dphi_{\rm B} \right| \nonumber \\
&=\frac 1{2\pi}\int d\dphi \sum_{\vec u_m} p(\vec u_m|\dphi) \left|\int e^{i\sysphase_{\rm B}}P(\sysphase_{\rm B}-\feedphase_1|\vec u_m)d\sysphase_{\rm B} \right| \nonumber \\
&= {\rm E}\left[ |\langle e^{i\sysphase_{\rm B}}\rangle|\right].
\end{align}
The E indicates an expectation value over the measurement results and $\dphi=\sysphase-\feedphase_1$. This is independent of the actual system phase, and may therefore be regarded as an average over the measurement results and $\feedphase_1$. As above the angle brackets indicate an average over the system phase in the Bayesian probability distribution, $\sysphase_{\rm B}$. In the main text we simply call this $\sysphase$, as we do not have the actual system phase and system phase for the Bayesian probability distribution in the same expression.

\section{General measure of resources}
\label{sec:meas}
As explained in Sec.~II, we adopt the general definition of $N$ in a system-independent way as the size of the smallest interval supporting the Fourier transform of the family of states $\ket{\psi(\phi)}$. Here 
$\ket{\psi(\phi)}$ is interpreted as follows. The entire measurement scheme may be regarded as preparation of a pure quantum state, followed by a unitary operation that depends on $\phi$, and finally a measurement. The preparation of the state and the $\phi$-dependent unitary operation may be grouped together into preparation of the $\phi$-dependent state $\ket{\psi(\phi)}$. In general there will be restrictions on both the input state (for example, limited photon number) and the unitary operation (for example, a limited number of passes through the phase shift) that will limit the possible states $\ket{\psi(\phi)}$.

Taking the Fourier transform of the family of states $\ket{\psi(\phi)}$ yields
\begin{equation}
\ket{\tilde\psi(\action)} = \frac{1}{\sqrt{2\pi}}\int_{-\infty}^{\infty} e^{-i\action\phi}\ket{\psi(\phi)} d\phi.
\end{equation}
The states $\ket{\tilde\psi(\action)}$ are not (necessarily) normalized. We consider the case where  restrictions on the measurement scheme imply that the family $\ket{\tilde\psi(\action)}$ has support on a finite interval $[N_{\rm min},N_{\rm max}]$, and so   may be represented as
\begin{equation}
\label{eq:freqs}
\ket{\psi(\phi)} = \frac{1}{\sqrt{2\pi}}\int_{N_{\rm min}}^{N_{\rm max}} e^{i\action\phi}\ket{\tilde\psi(\action)} d\action,
\end{equation}
We define $N$ as the smallest value of $N_{\rm max}-N_{\rm min}$.

There are two main cases that we can consider: 
\begin{enumerate}
\item $\ket{\tilde\psi(\action)}$ has support on a discrete set $\cu{s_j} \in [N_{\rm min},N_{\rm max}]$,
with every $s_i-s_j$ being integer-valued,
\item $\ket{\tilde\psi(\action)}$ may be nonzero for any $\action \in [N_{\rm min},N_{\rm max}]$.
\end{enumerate}
In the first case, $\ket{\psi(\phi)}$ is periodic in $\phi$ with period $2\pi$. Then $\phi$ is a phase, and the variance may be measured via the Holevo variance. In the second case there is no periodicity imposed on $\ket{\psi(\phi)}$. The appropriate measure of the variance of $\phi$ is then the usual variance. 

The first thing to notice is that the states $\ket{\tilde\psi(\action)}$ may be taken to be orthogonal without affecting the minimum variance that may be obtained. This can be proven as follows. Writing these states in some basis of orthogonal states $\ket{\zeta}$,
\begin{equation}
\label{eq:expand}
\ket{\psi(\phi)} = \frac{1}{\sqrt{2\pi}}\int_{N_{\rm min}}^{N_{\rm max}} e^{i\action\phi}\int_{-\infty}^{\infty} \psi(\action,\zeta)\ket{\zeta} d\zeta d\action.
\end{equation}
Now we can construct the modified state $\ket{\psi'(\phi)}$ by adding an additional subspace with orthogonal states $\ket{\action}$.
\begin{equation}
\ket{\psi'(\phi)} = \frac{1}{\sqrt{2\pi}}\int_{N_{\rm min}}^{N_{\rm max}} e^{i\action\phi} \ket{\action} \int_{-\infty}^{\infty} \psi(\action,\zeta)\ket{\zeta}d\zeta d\action.
\end{equation}
This modified state is not necessarily normalized.
Performing a measurement of this new subspace via projections $\ket{\theta}\bra{\theta}$, with
\begin{equation}
\ket{\theta} = \frac{1}{\sqrt{2\pi}}\int_{-\infty}^{\infty} e^{i\action\theta} \ket{\action} d\action,
\end{equation}
yields
\begin{equation}
\braket{\theta}{\psi'(\phi)} = \frac{1}{2\pi}\int_{N_{\rm min}}^{N_{\rm max}} e^{i\action(\phi-\theta)} \int_{-\infty}^{\infty} \psi(\action,\zeta)\ket{\zeta}d\zeta d\action.
\end{equation}
This state is unchanged from that in Eq.\ \eqref{eq:expand}, except for a shift in $\phi$ of $\theta$. Thus the state $\ket{\psi'(\phi)}$ can not give any less information about $\phi$ than $\ket{\psi(\phi)}$. In addition, the state $\ket{\psi'(\phi)}$ is of the form \eqref{eq:freqs}, except $\ket{\tilde\psi(\action)}$ has been replaced with the orthogonal states
\begin{equation}
\ket{\action} \int_{-\infty}^{\infty} \psi(\action,\zeta)\ket{\zeta}d\zeta.
\end{equation}
Hence the states $\ket{\tilde\psi(\action)}$ may be replaced with orthogonal states with no loss of phase information. Thus, in examining the limits to estimation of $\phi$, we can assume the $\ket{\tilde\psi(\action)}$ are orthogonal with no loss of generality.

In the case where the frequencies $\cu{s_j}$ are restricted to differ by integer values, then $\ket{\tilde\psi(\action)}$ becomes a series of delta functions in $\action$. The integral  \eqref{eq:freqs}  becomes a sum
\begin{equation}
\ket{\psi(\phi)} = e^{iN_{\rm min}\phi}\sum_{\action=0}^{N} e^{i\action\phi}\tilde{\psi}(s)\ket{\action},
\end{equation}
where
\begin{equation}
\tilde{\psi}(s)\ket{\action} = \frac{1}{\sqrt{2\pi}} \int_{\action-\epsilon}^{\action+\epsilon} \ket{\tilde\psi(\action')} d\action',
\end{equation}
for $1>\epsilon>0$. Assuming the $\ket{\tilde\psi(\action)}$ are orthogonal, the $\ket{\action}$ are orthogonal as well. This state is then identical to that for a phase shift on a single mode with a maximum of $N$ photons, and it has a phase uncertainty lower bounded by Eq.\ \eqref{eq:minv}, which we repeat here:
\beq
\Delta\phi_{\rm HL} = \sqrt{V_{\rm H}} = \tan\left(\frac\pi{N+2}\right) \sim \frac{\pi}{N}.
\end{equation}

In the case where $\ket{\tilde\psi(\action)}$ is nonzero for any value of the frequency $\action \in [N_{\rm min},N_{\rm max}]$, we may find the bound on the variance as follows. Without loss of generality, the family of states can be given by 
\beq
\label{eq:conts}
\ket{\psi(\phi)} = \int_0^N e^{i\action\sysphase} \tilde{\psi}(s) \ket{s},
\eeq
where here we have normalized $\cu{\ket{s}}$ as $\braket{s}{s'}=\delta(s-s')$. In this continuous case the system is equivalent to position and momentum, with the frequency $s$ equivalent to position, and $\phi$ equivalent to momentum (provided we use units where $\hbar=1$). It is therefore clear that the optimal measurement for $\phi$ is a projection in the basis
\begin{equation}
\ket{\phi} = \int_{-\infty}^{\infty} e^{is\phi}\ket{s}.
\end{equation}
This yields a phase variance of
\beq
V = \int_{-\infty}^{\infty} \phi^2 |\psi(\phi)|^2 d\phi - \ro{\int_{-\infty}^{\infty} \phi |\psi(\phi)|^2 d\phi}^2,
\eeq
where $\psi(\phi)$ is the inverse Fourier transform of $\tilde{\psi}(s)$. This result is for  $\sysphase$ initially completely unknown, with a flat prior distribution over the whole real line.

If we apply a phase shift of $\Delta\phi$ to the state, then this simply changes the mean value of $\phi$, but leaves the variance unchanged. Therefore, in considering the problem of minimizing the variance, we may consider states with mean $\phi$ equal to zero without loss of generality. The minimum uncertainty state then follows from the analogy of position and momentum. The bounds of 0 and $N$ on $s$ are equivalent to an infinite square well potential of size $N$. The minimum energy of a state in this infinite square well is $E=\pi^2/(2N^2)$, where we have taken $\hbar=1$ and the mass equal to 1. Because $\phi$ is equivalent to the momentum, and we are taking the mean $\phi$ to be equal to zero, $V=2E$, so the minimum variance is $V=\pi^2/N^2$. This gives
the lower bound on the uncertainty of 
\beq
\Delta\phi_{\rm HL} = \sqrt{V } =   \frac{\pi}{N}.
\eeq 
This result may also be found by taking the limit of the discrete case.

The case where the power of $e^{i\phi}$ can only take integer values is relevant to interferometry and to the gate formalism of Ref.\ \cite{vanDam2007}. For photons, each pass of a single photon through the phase shift gives a multiplication by a factor of $e^{i\phi}$, so $N_{\rm max}$ corresponds to an upper limit on the number of photon passes for the measurement scheme. It is not possible to have a negative number of photon passes, and normally there is no lower bound on the number of photon passes, so $N_{\rm min}=0$ and $N$ is the maximum number of photon passes. In cases where there is a minimum number of photon passes, then $N$ can also be taken to be the difference between the minimum and maximum numbers of photon passes, and yield a tighter bound on the phase uncertainty.

In the gate formalism of Ref.\ \cite{vanDam2007} each application of the phases shift is a gate of the form
\begin{equation}
\left[ {\begin{array}{*{20}c}
   1 & 0  \\
   0 & {e^{i\phi } }  \\
\end{array}} \right],
\end{equation}
applied to a qubit. Each application gives multiplication by a factor of $e^{i\phi}$ for the $\ket{1}$ basis state, and no multiplication for the $\ket{0}$ basis state. Therefore an upper bound on the power of $e^{i\phi}$, $N_{\rm max}$, is given by the number of applications of the phase shift. The lower bound is zero, so $N$ is simply the number of applications of the phase shift. In particular cases the details of the measurement scheme may mean that there are tighter bounds on the power. For example, for some gates the qubit may be initialized to $\ket{1}$, in which case $N_{\rm min}$ could be taken to be greater than zero. As in the case of photons, this would yield a tighter bound on the phase uncertainty. 

The third application we discuss is that of metrology of a Hamiltonian with an unknown parameter $\phi$,   describable as $H_\phi = \phi H$.  We consider estimation of $\phi$ by probing the Hamiltonian for total time $T$. This is the total time that systems evolve under the Hamiltonian, which can include parallel or serial evolution of systems. The maximum power of $e^{i\phi}$ that can be obtained is $-\lambda_{\rm min}T$, where $\lambda_{\rm min}$ is the minimum eigenvalue of $H$. The minimum power that can be obtained is $-\lambda_{\rm max}T$, where $\lambda_{\rm max}$ is the maximum eigenvalue of $H$. The difference between the upper and lower bounds on the power of $e^{i\phi}$ is therefore $N=\|H\|T$, where $\|H\|$ is the difference between the maximum and minimum eigenvalues of $H$.

In the case of metrology of a Hamiltonian, the time intervals can be any real numbers, so the powers of $e^{i\phi}$ can take any real values within the bounds. Therefore, $\phi$ may be measured as any real number, rather than modulo $2\pi$, and the bound on the phase uncertainty is exactly $\pi/N=\pi/(\|H\|T)$, and is measured by the standard deviation. In Ref.\ \cite{Caves08} the total probe time is $\nu t$, because there are $\nu$ independent probes of time $t$. The limit to phase measurement precision there is again $\pi/N$, with $N=\nu t\|H\|$. Note that Ref.\ \cite{Caves08} uses the terminology ``fundamental limit'' for the minimum possible phase uncertainty, and uses the terminology ``Heisenberg limit'' in a different sense.

\end{document}